%% file: 0-Main.tex
\documentclass[sigconf, nonacm]{acmart}

\usepackage{multirow}
\usepackage[ruled,vlined]{algorithm2e}
\usepackage{enumitem}
\usepackage{balance}
\usepackage{subfigure}
\usepackage{color}
\usepackage{colortbl}
\usepackage{soul}

\newtheorem{definition}{Definition}

\newtheorem{example}{Example}
\newtheorem{lemma}{Lemma}
\newtheorem{theorem}{Theorem}

\newcommand{\AlgoName}[1]{\textsf{#1}}
\newcommand{\FuncName}[1]{\textsf{#1}}
\newcommand{\ProbName}[1]{\textit{#1}}
\newcommand{\HidProf}[1]{#1}
\usepackage[show]{boxnotes}

\newcommand{\rev}[1]{#1}

\newcommand{\FullVersion}[1]{#1}
\newcommand{\PubVersion}[1]{}

% ...

\makeatletter
\patchcmd{\@algocf@start}
  {-1.5em}
  {0pt}
  {}{}
\makeatother

\newcommand\vldbdoi{10.14778/3636218.3636223}
\newcommand\vldbpages{657 - 670}
\newcommand\vldbvolume{17}
\newcommand\vldbissue{4}
\newcommand\vldbyear{2023}
\newcommand\vldbauthors{\authors}
\newcommand\vldbtitle{\shorttitle} 
\newcommand\vldbavailabilityurl{https://github.com/ZJU-DAILY/TBFC}
\newcommand\vldbpagestyle{empty} 

\begin{document}
\title{Efficient Temporal Butterfly Counting and Enumeration on Temporal Bipartite Graphs}

\author{Xinwei Cai}
\affiliation{%
  \institution{Zhejiang University}
}
\email{xwcai98@zju.edu.cn}

\author{Xiangyu Ke}
\authornote{Xiangyu Ke is the corresponding author.}
\affiliation{%
  \institution{Zhejiang University}
}
\email{xiangyu.ke@zju.edu.cn}

\author{Kai Wang}
\affiliation{
  \institution{ACEM, Shanghai Jiao Tong University} 
}
\email{w.kai@sjtu.edu.cn}

\author{Lu Chen}
\affiliation{
  \institution{Zhejiang University}
 }
\email{luchen@zju.edu.cn}

\author{Tianming Zhang}
\affiliation{
  \institution{Zhejiang University of Technology}
}
\email{tmzhang@zjut.edu.cn}

\author{Qing Liu}
\affiliation{
  \institution{Zhejiang University}
}
\email{qingliucs@zju.edu.cn}

\author{Yunjun Gao}
\affiliation{
  \institution{Zhejiang University}
}
\email{gaoyj@zju.edu.cn}

\begin{abstract}
Bipartite graphs characterize relationships between two different sets of entities, like actor-movie, user-item, and author-paper. The butterfly, a 4-vertices 4-edges (2,2)-biclique, is the simplest cohesive motif in a bipartite graph and is the fundamental component of higher-order substructures. Counting and enumerating the butterflies offer significant benefits across various applications, including fraud detection, graph embedding, and community search. While the corresponding motif, the triangle, in the unipartite graphs has been widely studied in both static and temporal settings, the extension of butterfly to temporal bipartite graphs remains unexplored.
In this paper, we investigate the {\em temporal butterfly counting and enumeration} problem: count and enumerate the butterflies whose edges establish following a certain order within a given duration. Towards efficient computation, we devise a non-trivial baseline rooted in the state-of-the-art butterfly counting algorithm on static graphs, further, explore the intrinsic property of the temporal butterfly, \rev{and develop a new optimization framework with a compact data structure and effective priority strategy.} The time complexity is proved to be significantly reduced without compromising on space efficiency. In addition, we generalize our algorithms to practical streaming settings and multi-core computing architectures. Our extensive experiments on 11 large-scale real-world datasets demonstrate the efficiency and scalability of our solutions.
\end{abstract}

\maketitle

\pagestyle{\vldbpagestyle}
\begingroup\small\noindent\raggedright\textbf{PVLDB Reference Format:}\\
\vldbauthors. \vldbtitle. PVLDB, \vldbvolume(\vldbissue): \vldbpages, \vldbyear.\\
\href{https://doi.org/\vldbdoi}{doi:\vldbdoi}
\endgroup
\begingroup
\renewcommand\thefootnote{}\footnote{\noindent
This work is licensed under the Creative Commons BY-NC-ND 4.0 International License. Visit \url{https://creativecommons.org/licenses/by-nc-nd/4.0/} to view a copy of this license. For any use beyond those covered by this license, obtain permission by emailing \href{mailto:info@vldb.org}{info@vldb.org}. Copyright is held by the owner/author(s). Publication rights licensed to the VLDB Endowment. \\
\raggedright Proceedings of the VLDB Endowment, Vol. \vldbvolume, No. \vldbissue\ %
ISSN 2150-8097. \\
\href{https://doi.org/\vldbdoi}{doi:\vldbdoi} \\
}\addtocounter{footnote}{-1}\endgroup

\ifdefempty{\vldbavailabilityurl}{}{
\vspace{.3cm}
\begingroup\small\noindent\raggedright\textbf{PVLDB Artifact Availability:}\\
The source code, data, and/or other artifacts have been made available at \url{\vldbavailabilityurl}.
\endgroup
}

\input{1-Introduction}
\input{2-Preliminaries}
\input{3-Baseline}
\input{4-Opt}
\input{5-Streaming}
\input{6-Exp}
\input{7-RelatedWork}
\input{8-Conclusion}

\begin{acks}
This work was supported in part by the NSFC under Grants No. (62025206, U23A20296, 62302444, 62302294, and 62302451) and NSF of Zhejiang under Grant No. LQ22F020018.
\end{acks}

\newpage
\balance
\clearpage

\bibliographystyle{ACM-Reference-Format}
\bibliography{ref}

\FullVersion{\input{9-Approximation}}

\end{document}

%% file: 1-Introduction.tex
\section{Introduction}
\label{sec:intro}

Bipartite graphs, which separate vertices into two disjoint sets and allow edges only between different sets of vertices, serve as natural data models for capturing relationships between two distinct types of entities~\cite{zhou2007bipartite}, such as actor-movie, user-item, and author-paper. As motifs (i.e., small frequent subgraph patterns) are fundamental building blocks of complex graphs~\cite{seshadhri2013triadic}, discovering and counting motifs can reveal hidden relationships among participating entities~\cite{ahmed2017graphlet, bressan2017counting, seshadhri2013triadic}, contributing to the characterization of complex networks~\cite{milo2002network}, such as social network analysis~\cite{yang2018node}, traffic speed forecasting~\cite{zhang2019hybrid}, and research on spiking activity in neural networks~\cite{hu2013motif}. In the context of bipartite graphs, butterfly (i.e. a (2,2)-biclique) - the simplest cohesive higher-order sub-structure, is the most fundamental motif, analogous to the triangle in unipartite graphs~\cite{aksoy2017measuring}. A bipartite graph cannot exhibit any community structure without butterflies as analyzed in~\cite{aksoy2017measuring}. Counting and enumerating butterflies have become essential components in various downstream network analytic tasks, e.g., bipartite clustering coefficient computation~\cite{aksoy2017measuring}, $k$-bitruss construction~\cite{sariyuce2018peeling}, graph embedding~\cite{huang2021signed}, etc.

\begin{figure}[tbp]
	\centering
	\includegraphics[width=2.7in]{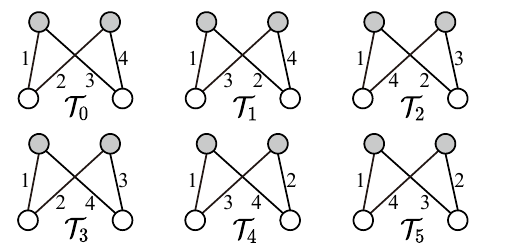}
	\vspace{-0.12in}
	\caption{6 types of non-isomorphic temporal butterfly. The vertices from the same layer have the same color (grey in $U$ and white in $L$), and the number denotes the temporal order.}
	\label{fig:tbf_type}
	\vspace*{-0.23in}
\end{figure}

In real-world scenarios, networks exhibit a temporal nature, where interactions between entities can arise and cease over time. To capture such dynamics, temporal graphs are employed, where edges are associated with timestamps~\cite{liu2021temporal, kovanen2011temporal}. \rev{By incorporating the temporal ordering and duration constraint (i.e., all edges have to occur within a fixed duration)~\cite{paranjape2017motifs, gao2022scalable}, temporal motifs offer enhanced information and greater expressiveness compared to standard motifs. Temporal ordering represents the sequence of events, while temporal duration denotes the validity period of these events.} Whilst \ProbName{temporal motif counting and enumeration} have been extensively studied on temporal unipartite graphs~\cite{kovanen2011temporal, paranjape2017motifs, gao2022scalable, pashanasangi2021faster, LociceroMPF20, mackey2018chronological}, the temporal bipartite graphs are yet to be explored, except for fundamental reachability query \cite{chen2021efficiently}. 
Motivated by this research gap, we investigate the \ProbName{temporal butterfly counting and enumeration} problem, which is to count and enumerate butterflies in different temporal ordering (i.e., 6 non-isomorphic temporal butterfly types as shown in Figure~\ref{fig:tbf_type}) while adhering to duration constraint.

\noindent\textbf{Applications.} We present two representative real-world applications of temporal butterflies as below.

\begin{figure}[bp]
	\vspace*{-0.32in}
	\centering
	\includegraphics[width=2.8in]{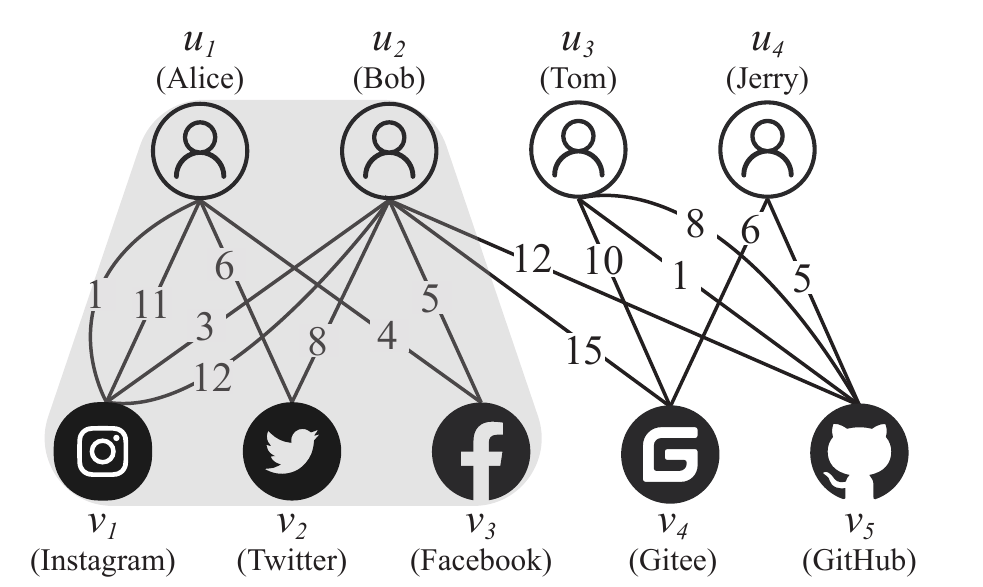}
	\vspace{-0.1in}
	\caption{A user-website network.}
	\label{fig:example}
\end{figure}

\textit{[Recommendation]}: When providing recommendations on a user-item network, the edges of the network are naturally timestamped to indicate the timing of interactions. For example, in the context of a movie recommendation system, the timestamped edges can represent instances when a customer watches a movie. Similar scenarios may include paper reading or user website exploration ~\cite{he2016birank, wang2006unifying, wang2020efficient, wang2022towards}. While static butterflies (encompass all six types in Figure~\ref{fig:tbf_type}) can capture user/item pairs with similar behavior, butterflies with specific temporal ordering can contribute to role recognition ~\cite{paranjape2017motifs}. \rev{For instance, types $\mathcal{T}_0$, $\mathcal{T}_1$, and $\mathcal{T}_2$ (as shown in Figure~\ref{fig:tbf_type}) can indicate that a user consistently follows another user in their behavior when considering the grey vertices as users. In particular, type $\mathcal{T}_0$ captures the immediate co-doing behavior, suggesting a stronger follower effect. This temporal information is valuable for modeling social influence \cite{chen2021efficiently}. In Figure~\ref{fig:example}, the notably higher count of type $\mathcal{T}_0$ instances between Alice and Bob suggests that Bob often accesses the same website in close succession after Alice does. This pattern strongly implies the possibility that Bob might be Alice's follower.} Additionally, the duration constraint strengthens the intrinsic correlation between entities, enhancing the accuracy and effectiveness of recommendations~\cite{mackey2018chronological, li2019time, LociceroMPF20}. 

\textit{[Data Monitor]}: In decentralized finance, users' transaction information on each platform is publicly accessible, enabling the construction of a user-platform network~\cite{jensen2021introduction}. While transactions are secure and transparent, user anonymity poses challenges to effective monitoring. By incorporating temporal ordering into static butterflies, we can extract more accurate relationships between anonymous users. For instance, type $\mathcal{T}_3$ represents minimal asset circulation, while types $\mathcal{T}_4$ and $\mathcal{T}_5$ indicate asset exchanges between two users (assuming grey vertices represent users and white vertices represent platforms). Introducing a stricter duration constraint can enhance the monitoring of high-frequency transactions. 

Other applications include disease control in people-location network~\cite{eubank2004modelling}, fraud detection on user-page networks~\cite{liu2019efficient}, threat hunting in process-IP network~\cite{li2021threat}, etc.

\noindent\textbf{Challenges.} Although one can sacrifice the bipartite property and apply existing temporal motif counting and enumeration techniques to determine that of the 4-edge rectangle (then filter based on vertex type), existing techniques either fail since they only support up to 3-vertex, 3-edge temporal motifs~\cite{kovanen2011temporal, paranjape2017motifs, gao2022scalable, pashanasangi2021faster}, or become extremely inefficient\footnote{They cannot avoid permuting all possible combinations of four orderly edges within a duration constraint and check if it is a butterfly, which takes $O(|E|^4)$ time.} \cite{paranjape2017motifs, mackey2018chronological, li2019time}.
Moreover, existing butterfly counting and enumeration techniques cannot be easily adapted to the temporal environment, as they struggle with handling the more complicated situations on the temporal bipartite graph (i.e., there may be multiple edges between two vertices), and require additional overhead to check whether the identified butterflies meet the specified constraints and determine the types correctly.

\noindent\rev{\textbf{Our Contributions.} We are the first to study the \textit{temporal butterfly counting and enumeration} problem, which aims to quantify the number of butterflies or enumerate instances of butterflies whose edges follow a specific temporal ordering within a specified time duration (for precise details, please refer to \S~\ref{sec:pre} for the formal definition).}

Our baseline solutions build upon the state-of-the-art \ProbName{butterfly counting} algorithm \cite{WangLQZZ19} through refining the vertex-priority assignment, verifying the duration constraint, and casting all possible temporal ordering (\S~\ref{sec:baseline}). 
We carefully observe and summarize the rules governing the relationships between two wedges, which are the core components of a butterfly. Based on these observations, optimization techniques are designed accordingly  (\S~\ref{sec:framework}).
\rev{In particular, we devise the compact data structure (i.e., wedge set) and extend the priority concept from vertex to wedge level, which captures the temporal ordering from both direction and coverage perspectives (\S~\ref{sec:opt_overview}). The searching space is largely pruned and the redundant permutation is avoided. These optimization techniques enable efficient counting and enumeration algorithms with minor modifications to the technical framework, resulting in substantial gains in efficiency (\S~\ref{sec:opt}, \ref{sec:enum}). In addition, we incorporate advanced engineering efforts to handle extreme cases and further improve counting efficiency. Theoretically, the time complexity is significantly reduced while the space complexity remains unchanged (\S~\ref{sec:opt+}). Empirical evaluation over real-world datasets validates that our optimized algorithm is up to 3773.1 times faster than the baseline.}

\rev{To support the practical graph streams~\cite{wang2022efficient, pereira2016evolving, steer2020raphtory, miao2015immortalgraph}, we extend our counting algorithm to facilitate dynamic updates over streams and further propose a non-trivial parallel algorithm to handle batch updates (\S~\ref{sec:stream}). The parallel algorithm focuses on resolving the counting conflicts and providing problem-specific simplifications.}

Finally, we demonstrate the efficiency and scalability of our proposed algorithms via extensive experimental evaluations on 11 large-scale temporal bipartite networks.

Our principal contributions are summarised as follows.
\begin{itemize}
\vspace{-1mm}
	\item We are the first to define the concept of temporal butterflies and conduct a comprehensive study on the problem of \ProbName{temporal butterfly counting and enumeration} (\S~\ref{sec:pre}).
	
 	\item \rev{We design a non-trivial baseline based on the state-of-the-art \ProbName{butterfly counting} algorithm (\S~\ref{sec:baseline}) and develop an optimization framework with compact data structure and effective priority strategy.} Theoretically, the time complexity is significantly reduced without sacrificing the space (\S~\ref{sec:framework}).
	
 	\item We adapt our algorithm to temporal bipartite graph stream setting and further propose a parallel algorithm to improve the throughput of streaming data by batch counting (\S~\ref{sec:stream}).
	
 	\item We conduct extensive experiments on various real-world temporal bipartite graphs to demonstrate the efficiency and scalability of our proposed algorithms (\S~\ref{sec:exp}).
\end{itemize}

%% file: 2-Preliminaries.tex
\vspace{-0.06in}
\section{Preliminaries}
\vspace{-0.02in}
\label{sec:pre}




An undirected\footnote{Bipartite graphs are generally undirected in existing studies~\cite{WangLQZZ19, Sanei-MehriST18, zhou2007bipartite, aksoy2017measuring,sariyuce2018peeling} and real-world datasets (\S~\ref{sec:exp}).} {\em temporal bipartite graph} $G=(V=(U, L), E, T)$ is defined over two disjoint sets of vertices $U$ and $L$, i.e., $U \cap L = \emptyset, U \cup L = V$, which represent two classes of real-world entities, known as upper and lower layer vertex sets, respectively. The connections only exist across different classes, i.e., edge set $E \subseteq U \times L \times T$, where $T$ is a collection of timestamps. Each {\em temporal edge} $e=(u, v, t)\in E$ represents an interaction between $u$ and $v$ at the time $t$. Notice that multiple temporal edges may exist between the same pair of vertices with different timestamps. $E(u)$ denotes the set of temporal edges adjacent to vertex $u$. We extend the concept of {\em butterfly} and its basic component, {\em wedge}, from simple static graphs 
as follows: 

\begin{definition}[Temporal Wedge]\label{def:wedge}
    In a temporal bipartite graph $G$, a temporal wedge $\angle(u, v, w, t_s, t_a)$ is a 2-hop path consisting of $(u, v, t_s)$ and $(v, w, t_a)$. 
\end{definition}

The inherent nature of a bipartite graph ensures that $u$ and $w$ belong to one layer while $v$ is from the other layer. We designate $u$ the start-vertex, $v$ the middle-vertex, and $w$ the end-vertex. In the following discussions, a wedge is {\em forward} if $t_s < t_a$, and is {\em backward} if $t_s > t_a$.

\begin{definition}[Butterfly \cite{wang2014rectangle}]\label{def:bf}
    Given vertices $u, w \in U$ and $v, x \in L$, the subgraph induced by these four vertices in $G$ is a butterfly if it is a $2 \times 2$ bi-clique; that is, $u$ and $w$ are all connected to $v$ and $x$, respectively, by edges.
\end{definition}

A butterfly can be decomposed into a pair of wedges with the same start-vertex, the same end-vertex, and different middle-vertices. Therefore, most existing \ProbName{butterfly counting} algorithms~\cite{wang2014rectangle, sanei2018butterfly, WangLQZZ19} focus on efficient enumeration and permutation of wedges.

\begin{definition}[Temporal Butterfly]\label{def:tbf}
    Given a duration threshold $\delta$, a temporal butterfly is a sequence of 4 temporal edges in chronological order $\langle e_1, e_2, e_3, e_4 \rangle$, s.t., (1) $e_1.t<e_2.t<e_3.t<e_4.t$, (2) $e_4.t - e_1.t \le \delta$, and (3) the induced static graph is a butterfly.
\end{definition}

\rev{The duration constraint enforces that all edges must occur within a fixed duration $\delta$ to ensure the existence of a butterfly, i.e., the earliest edge has not yet expired when the latest edge appears. Regarding the graph depicted in Figure~\ref{fig:example}, upon setting $\delta$ to 15, we can identify two instances of type $\mathcal{T}_2$ involving the vertices $u_2, u_3, v_4, v_5$. However, if we impose a more stringent duration constraint of 10, only one butterfly will remain. The static butterfly fails to capture such kind of valuable property.}
The possible temporal permutations\footnote{Following \cite{paranjape2017motifs, kovanen2011temporal, pashanasangi2021faster}, we assume that the four timestamps on a temporal butterfly are distinct, which can be implemented by applying simple tie-breaking rule, e.g., based on the unique indexes of the starting/ending vertices as in \cite{WangLQZZ19}.} (i.e., 6 non-isomorphic temporal butterflies as shown in Figure~\ref{fig:tbf_type}) of these four edges make the induced butterfly even more expressive in real-world applications, as illustrated in \S~\ref{sec:intro}.

\noindent\textbf{Problem Statement}. Given a temporal bipartite graph $G$ and a threshold $\delta$, our \ProbName{temporal butterfly counting} problem is to count the number $\{C[i]\}_{i=0}^5$ of each of six types of temporal butterflies $\mathcal{T}_0, \mathcal{T}_1, \cdots, \mathcal{T}_5$, and our \ProbName{temporal butterfly enumeration} problem is to find all these temporal butterfly instances $\{B[i]\}_{i=0}^5$.

\noindent\textbf{Solution Overview}. \rev{We first devise our baseline solutions by extending the state-of-the-art static butterfly counting algorithm~\cite{WangLQZZ19} (\AlgoName{TBC}/\AlgoName{TBE} in \S~\ref{sec:baseline}). Then we optimize the algorithms by designing a compact data structure and generalizing priority to wedge level for smart pruning and processing acceleration (\AlgoName{TBC$^+$}/\AlgoName{TBE$^+$}/\AlgoName{TBC$^{++}$} in \S~\ref{sec:framework}). In addition, we generalize our counting solution to meet the practical steaming demand (\AlgoName{STBC}/\AlgoName{STBC$^+$} in \S~\ref{sec:stream}).}

%% file: 3-Baseline.tex
\vspace{-0.04in}
\section{Baseline Solution}
\vspace{-0.01in}
\label{sec:baseline}

In this section, we devise our baseline solution, based on the state-of-the-art \ProbName{butterfly counting} algorithm \AlgoName{BFC-VP}~\cite{WangLQZZ19}. Hereafter, we omit temporal in the temporal edge/wedge/butterfly when the context is clear.

\AlgoName{BFC-VP} sorts the vertices based on their proposed vertex priority and efficiently enumerates all but less redundant wedges that can form a butterfly. Following the same intuition, we assign a unique vertex priority for any vertex $u$ based on $|E(u)|$. Note that the priority is no longer neighbor-based as there may exist multiple temporal edges between two vertices. 
\PubVersion{\rev{The correctness and efficiency proofs follow~\cite{WangLQZZ19}, while details can be found in full version~\cite{cai2023efficient}.}}
\FullVersion{\rev{The correctness and efficiency proofs follow~\cite{WangLQZZ19}, while details can be found later.}}

\begin{definition}[Vertex Priority]\label{def:vp}
    For any vertex $u$ in a temporal bipartite graph $G$, the priority $P_V(u)$ is an integer in $[1, |V|]$. For any two vertices $u$ and $w$ in $G$, $P_V(u) > P_V(w)$ if: 
	\begin{itemize}
		\item $|E(u)| > |E(w)|$, or
		\item $|E(u)| = |E(w)|$ and $id(u) > id(w)$ 
	\end{itemize}
	where $id(u)$ is the unique vertex ID of $u$.
\end{definition}

\setlength{\textfloatsep}{0pt}
\begin{algorithm}[tbp]
	\caption{\AlgoName{TBC}}
	\label{algo:TBC}
	\linespread{0.8}\selectfont
	\LinesNumbered
	\DontPrintSemicolon
	\SetKwFunction{Type}{\FuncName{Type}}
	\SetKwFunction{IsTB}{\FuncName{IsTB}}
	\KwIn{the temporal bipartite graph $G=(V = (U, L), E, T)$; the threshold $\delta$}
	\KwOut{the counts $\{C[i]\}_{i=0}^5$}
	$\{C[i]\}_{i=0}^5 := 0$\;
	\ForEach{$E(u) : u \in V$}{
		sort all $e \in E(u)$ according to $P_V(e.v)$\;
	}
	\ForEach{$u \in V$}{
		initialize hashmap $H$ to store wedges\;
		\ForEach{$(u, v, t) \in E(u): P_V(u) > P_V(v)$}{
			\ForEach{$e'(v, w, t) \in E(v) : P_V(u) > P_V(w)$}{
				$H[w].$\FuncName{append}$((u, v, w, t, t'))$\;
			}
		}
		\ForEach{vertex $w \in H : |H[w]| > 1$}{
			\ForEach{{\bf pair} $\angle_i, \angle_j \in H[w] : j < i$}{ 
				\If{\IsTB{$\angle_i$, $\angle_j$}}{
					$C[$\Type{$\angle_i, \angle_j$}$]$ += $1$\;
				}
			}
		}
	}
	\Return{$\{C[i]\}_{i=0}^5$}
\end{algorithm}
\setlength{\textfloatsep}{12pt plus 2pt minus 2pt}

\AlgoName{TBC} follows a sequential process of ``enumerate-filter-match'', as outlined in Algorithm~\ref{algo:TBC}. After the initialization and priority assignment (line 1-3), \AlgoName{TBC} constructs the wedges from each start-vertex $u$ to all lower-priority vertices (line 4-8). Subsequently, for each possible wedge combination (line 9-10), \AlgoName{TBC} filters out invalid instances (line 11) and determines their type according to Figure~\ref{fig:tbf_type} (line 12). In particular, \FuncName{Type}() returns the butterfly type and \FuncName{IsTB}() (\underline{Is} \underline{T}emporal \underline{B}utterfly) checks for the following constraints: {\bf (1)} Middle-vertices of the two wedges should be different. {\bf (2)} There exists a temporal ordering for the four timestamps of the two wedges, i.e., no two timestamps are equal. {\bf (3)} All timestamps must fall within a $\delta$ duration, i.e., the difference between the maximum and minimum timestamps cannot exceed $\delta$. Additionally, we can obtain \AlgoName{TBE} by simply modifying Algorithm~\ref{algo:TBC} from counting to storing/outputting all instances (line 12).

\FullVersion{
\begin{figure}[htbp]
    \centering
    \includegraphics[width=0.9\linewidth]{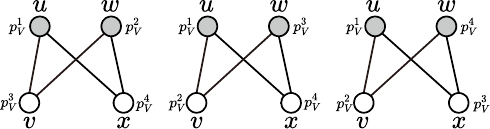}
    \vspace*{-0.1in}
    \caption{All possible static butterflies while $u$ is the start-vertex and the vertex priority follows $p_V^1 > p_V^2 > p_V^3 > p_V^4$.}
    \label{fig:correct}
    \vspace*{-0.1in}
\end{figure}
}
\FullVersion{\rev{
Below we show the correctness and the complexity of \AlgoName{TBC}.
\begin{theorem}
    The \AlgoName{TBC}  correctly solves the temporal butterfly counting problem.
\end{theorem}
}}
\FullVersion{\rev{
\textsc{Proof}. \AlgoName{TBC} follow the principle of identifying static butterflies, verifying their status as temporal butterflies, and subsequently classifying their type. A static butterfly can always be constructed from two distinct wedges that share the same start- and end-vertex but differ in their middle-vertices. Therefore, our objective is to formally demonstrate that each static butterfly within graph $G$ is correctly counted exactly once by \AlgoName{TBC}. Consider a butterfly formed by four vertices: $x, u, v, w$. Without loss of generality, let's assume that $u$ is designated as the start-vertex and holds the highest vertex priority. The distribution of vertex priorities can fall into one of three scenarios, as depicted in Figure~\ref{fig:correct} (the other situations can be transformed into the above by a symmetric conversion), where $p_V^i$ represents the corresponding vertex priority.
Regarding the case in Figure~~\ref{fig:correct}, \AlgoName{TBC} only counts the static butterfly once from the wedges $(u, v, w)$ and $(u, x, w)$. 
The following-up status checking and bufferfly type determining only involve simple duration constraint verification and the mapping to all possible ordering, which will not introduce errors.
This substantiates our claim that \AlgoName{TBC} accurately addresses the \textit{temporal butterfly counting problem}. \hfill $\square$
}}

\noindent\textbf{Complexity Analysis}. 

\noindent$\bullet$ The time complexity of \AlgoName{TBC} is $O(\sum_{u\in V}|W(u)|^2)$, where \\$|W(u)|=\sum_{(u,v,t) \in E(u):|E(v)|\leq|E(u)|}|E(v)|$.

\textsc{Proof}. The first phase of \AlgoName{TBC} enumerates all possible wedge instances in $O(\sum_{(u,v) \in E}\FuncName{min}\{|E(u)|,|E(v)|\})$, as reported by \cite{WangLQZZ19} for \AlgoName{BFC-VP}. In the second phase of butterfly construction, \AlgoName{TBC} takes quadratic time overhead to check the various temporal conditions. We denote all the processed wedges with the start-vertex $u$ by $W(u)$, and $|W(u)|=\sum_{(u,v,t) \in E(u):|E(v)|\leq|E(u)|}|E(v)|$. Therefore, the overall time complexity of \AlgoName{TBC} is $O(\sum_{u\in V}|W(u)|^2)$. \hfill $\square$

\noindent$\bullet$ The space complexity of \AlgoName{TBC} is $O(|E|+\FuncName{max}_{u\in V}\{|W(u)|\})$, where $|W(u)|=\sum_{(u,v,t) \in E(u):|E(v)|\leq|E(u)|}|E(v)|$.

\textsc{Proof}. As demonstrated in the above proof for time complexity analysis, in addition to the input graph, \AlgoName{TBC} only stores the wedges of one particular start-vertex (and discards them after the current iteration). The space complexity simply follows. \hfill $\square$

\noindent$\bullet$ The time/space complexity of \AlgoName{TBE} is the same as \AlgoName{TBC}.

\textsc{Proof}. Assuming that all butterfly instances are directly output to disk without occupying memory, there is no essential difference between \AlgoName{TBC} and \AlgoName{TBE}. \hfill $\square$

%% file: 4-Opt.tex
\vspace{-0.06in}
\section{A New Framework with Wedge Set}
\vspace{-0.02in}
\label{sec:framework}

In this section, we first discuss the intuitions behind our optimizations (specifically for counting), inspired by the observations of butterfly types. Then, we present our detailed counting algorithm designs and smart implementations. With minor modifications, the algorithm can also be adapted for enumeration purposes. Finally, we apply advanced data structures to enhance our counting efficiency and effectively handle extreme cases.

\vspace{-0.06in}
\subsection{Optimization Overview}
\vspace{-0.02in}
\label{sec:opt_overview}

\begin{figure}[htbp]
	\vspace*{-0.12in}
	\centering
	\includegraphics[width=3.0in]{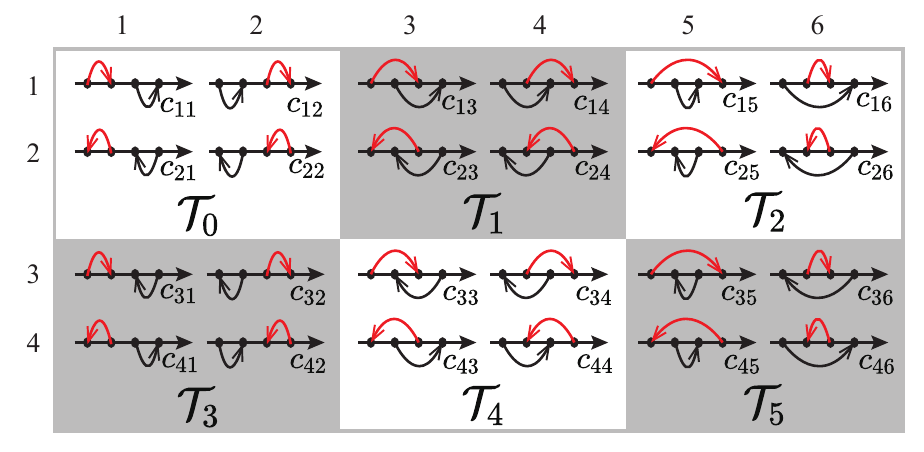}
	\vspace{-0.1in}
	\caption{\rev{All possible temporal orderings of two temporal wedges $\angle_i$ and $\angle_j$, represented by red and black arcs above/under a time axis. The arcs point from $t_s$ to $t_a$.}}
	\label{fig:tbf_case}
	\vspace*{-0.10in}
\end{figure}

Figure~\ref{fig:tbf_case} illustrates all 24 potential temporal orderings between two arbitrary temporal wedges, denoted as $\angle_i$ and $\angle_j$. These wedges share the same start-vertex and end-vertex in $U$ but differ in their middle-vertex in $L$. Upon analyzing the temporal relations between the time arcs induced by $\angle_i$ and $\angle_j$, we make the following observations:
{\bf (1)} From the {\em temporal coverage} perspective, there are three possible categories: {\em non-overlap}, {\em intersecting}, and {\em covering}. 
{\bf (2)} From the {\em temporal direction} perspective, they can either follow the same ({\em forward} or {\em backward}) direction or deviate from each other. The columns in \rev{Figure~\ref{fig:tbf_case}} correspond to the coverage relationship \{{\em non-overlap}, {\em intersecting}, and {\em covering} \} (from left to right), while the rows indicate whether the two wedges follow the same temporal direction or not. Each temporal ordering ($c_{xy}$ for that in the $x^{th}$ row and the $y^{th}$ column of Figure~\ref{fig:tbf_case}) in every distinct box can be transformed into one another by exchanging the wedge indices and/or reversing the start- and end-vertex, corresponding to one temporal butterfly type. In summary, the temporal butterfly types are built on 3 coverage patterns and 2 direction patterns. \rev{Therefore, we devise a compact data structure (i.e., wedge set) to distinguish the temporal directions and generalize the priority concept to the wedge level for smart pruning. In addition, we introduce the type conversion rule for counting the butterflies from either vertex layer.}

\noindent\textbf{Wedge Set}. 
In each iteration, we enumerate wedges starting from a specific start-vertex. In the presence of multiple edges, it is possible to have multiple wedges with the same middle-vertex. \rev{To optimize the process, we propose organizing the wedges into different sets based on their middle-vertices, as defined in Definition~\ref{def:set}. By doing so, during butterfly construction, we only consider wedges from different sets, effectively reducing time overhead. Furthermore, this approach reduces space requirements during counting, as we only need to store two timestamps for each wedge.}

\rev{Additionally, we partition the wedge set into two disjoint subsets, namely $A$ and $D$, for forward and backward wedges respectively.} To accommodate backward wedges, we swap the $t_s$ and $t_a$ values before inserting them into subset $D$. This simple operation significantly reduces the 24 possible orderings depicted in Figure~\ref{fig:tbf_case} to just 6 cases, as shown in the first row where each column becomes a merged case. However, despite the merging, determining the butterfly type remains straightforward: $A\times A$ or $D\times D$ indicates the same temporal direction ($\mathcal{T}_0$, $\mathcal{T}_1$, $\mathcal{T}_2$), while $A\times D$ or $D\times A$ represents different temporal directions ($\mathcal{T}_3$, $\mathcal{T}_4$, $\mathcal{T}_5$).

\begin{definition}[Wedge Set]\label{def:set}
    For all the wedges having the same start-vertex and the same end-vertex, those wedges with the same middle-vertex $v$ are stored in the wedge set $S_v=(A,D)$. For any such wedge, if its $t_s<t_a$, $(t_s, t_a)$ is stored in subset $A$, otherwise $(t_a,t_s)$ is stored in subset $D$. $A \cap D = \emptyset$.
\end{definition}

\noindent\textbf{Wedge Priority}. To avoid redundant permutations and facilitate further optimization, we prioritize the wedges when considering two arbitrary temporal wedges $\angle_i$ and $\angle_j$ that share the same start-vertex and end-vertex but differ in their middle-vertex. \rev{By constructing the butterflies in a wedge-priority-increasing manner, we effectively eliminate the need to handle cases $c_{12}$, $c_{14}$, and $c_{16}$. Instead, we can focus on determining the type of a temporal butterfly by examining the way the wedge sets join and evaluating three coverage patterns.} In cases where two wedges have the same wedge priority, their order can be arbitrarily chosen without affecting the correctness of subsequent algorithms.

\begin{definition}[Wedge Priority]\label{def:sp}
    The wedge priority $P_W(\cdot)$ is a total order among all wedges. For any two wedges $\angle_i$ and $\angle_j$, $P_W(\angle_i) < P_W(\angle_j)$ if:
	\begin{itemize}
		\item $\angle_i.t_s > \angle_j.t_s$, or
		\item $\angle_i.t_s = \angle_j.t_s$ and $\angle_i.t_a < \angle_j.t_a$.
	\end{itemize}
\end{definition}

\begin{figure}[htbp]
	\vspace*{-0.1in}
	\centering
	\includegraphics[width=1.8in]{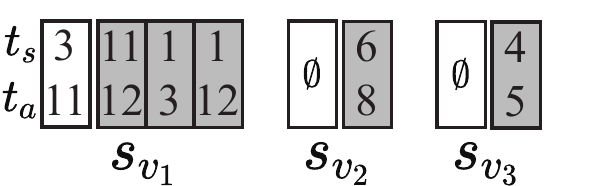}
	\vspace{-0.1in}
	\caption{The wedge sets construct from Figure~\ref{fig:example} while $u_2$ is the start-vertex and $u_1$ is the end-vertex.}
	\label{fig:set}
	\vspace*{-0.1in}
\end{figure}

\begin{example}
	Figure~\ref{fig:set} shows the 3 wedge sets of Figure~\ref{fig:example}, where $u_2$ is the start-vertex and $u_1$ is the end-vertex (since $P_V(u_2) > P_V(u_1)$). Each small rectangle with two numbers represents a wedge with two timestamps $t_s, t_a$, where the white ones are in $A$ and the gray ones are in $D$, and each subset is already sorted according to wedge priority.
\end{example}

\begin{figure}[htbp]
	\vspace*{-0.1in}
	\centering
	\includegraphics[width=3.0in]{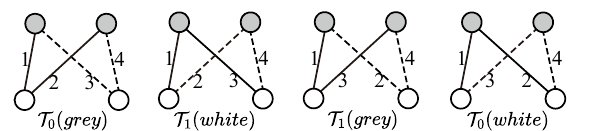}
	\vspace{-0.1in}
	\caption{Example for wedge type conversion. Two wedges are marked by solid and dashed lines, respectively.}
	\label{fig:conversion}
	\vspace*{-0.1in}
\end{figure}

\noindent\textbf{Type Conversion}. In Figure~\ref{fig:tbf_case}, we only discuss the cases while the start-vertex is in $U$, but both layers can serve as the starting side in practice\footnote{In real-world scenarios, our interest goes beyond examining the co-behavior of users. We also seek to understand the relationships between different items.}. \rev{Depending on the perspective of different layers, the same butterfly can be decomposed into different sets of wedges, resulting in distinct coverage and direction patterns.} For example, if the grey vertices are in $U$, then the butterfly belongs to type $\mathcal{T}_0$; otherwise, it falls into type $\mathcal{T}_1$. This relationship holds vice versa for the two butterflies on the right-hand side of Figure~\ref{fig:conversion}. Similar patterns can be observed between $\mathcal{T}_2$ and $\mathcal{T}_3$, as well as $\mathcal{T}_4$ and $\mathcal{T}_5$. \rev{Therefore, we can handle wedge combinations in a unified framework, and finally decide the butterfly type based on the conversion rule.}

\vspace{-0.08in}
\subsection{Algorithm Design}
\label{sec:opt}
\rev{The introduction of our optimization algorithm, \AlgoName{TBC$^+$}, will be structured in a gradual manner, progressing from an overarching perspective (specifically, Algorithm~\ref{algo:TBC+}) down to the finer intricacies.} \AlgoName{TBC$^+$} shares the same initialization and vertex priority assignment as \AlgoName{TBC} (lines 1-3). For each start-vertex $u$, \AlgoName{TBC$^+$} initializes a nested hashmap $H$ to store the wedges induced by each pair of end-vertex $w$ and middle-vertex $v$ (lines 4-8). This hashmap contains only two timestamps for each wedge and is organized based on the temporal directions (lines 9-12). Furthermore, non-empty sets with different middle-vertices are re-indexed, and the \FuncName{Combine}() function is used to compute the number of temporal butterflies (lines 13-16), which will be explained in detail in the following paragraphs. To optimize the algorithm, we employ simple pruning by excluding wedges with $|t_s - t_a| > \delta \vee t_s = t_a$. \rev{This pruning step performs a partial check of the temporal duration constraint in advance, filtering out illegal wedges. It ensures that each wedge $\angle$ satisfies $\angle.t_s < \angle.t_a \le \angle.t_s + \delta$.}

\begin{lemma}\label{lem:pruning}
	Given a temporal threshold $\delta$, a temporal wedge $\angle(t_s, t_a)$ with $|t_s - t_a| > \delta \vee t_s = t_a$ cannot be a part of any temporal butterfly.
\end{lemma}

\FullVersion{\rev{\textsc{Proof}. Given the temporal ordering and the temporal duration constraint of temporal butterflies, this can be proven trivially. \hfill $\square$}}

\PubVersion{\rev{This proof, along with other straightforward proofs, is immediate and is relocated to the full version~\cite{cai2023efficient}.}} 

\setlength{\textfloatsep}{0pt}
\begin{algorithm}[tbp]
	\caption{\AlgoName{TBC$^+$}}
	\label{algo:TBC+}
	\linespread{0.8}\selectfont
	\LinesNumbered
	\DontPrintSemicolon
	\SetKwFunction{Combine}{\FuncName{Combine}}
	\KwIn{the temporal bipartite graph $G=(V = (U, L), E, T)$; the threshold $\delta$}
	\KwOut{the counts $\{C[i]\}_{i=0}^5$}
	$\{C[i]\}_{i=0}^5 := 0$\;
	\ForEach{$E(u) : u \in V$}{
		sort all $e \in E(u)$ according to $P_V(e.v)$\;
	}
	\ForEach{$u \in V$}{
		initialize hashmap $H$ to store sets\;
		\ForEach{$(u, v, t) \in E(u): P_V(u) > P_V(v)$}{
			\ForEach{$(v, w, t') \in E(v) : P_V(u) > P_V(w)$}{
				\If{$t \neq t' \wedge |t' - t| \leq \delta$}{
					\uIf{$t < t'$}{
						$H[w][v].A.$\FuncName{append}$(\angle(t,t'))$\;
					}
					\ElseIf{$t > t'$}{
						\tcp*[l]{$t, t'$ is swaped when append}
						$H[w][v].D.$\FuncName{append}$(\angle(t',t))$\;
					}
				}
			}
		}
		\ForEach{vertex $w \in H : |H[w]| > 1$}{
			reindex sets in $H[w]$ to $S_0, S_1, \cdots, S_{|H(w)|}$\;
			sort each subset in $H[w]$ according to $P_W(\angle)$\;
			\rev{\Combine{$u, \delta, H[w], \{C[i]\}_{i=0}^5$}}\;
		}
	}
	\Return{$\{C[i]\}_{i=0}^5$}
\end{algorithm}
\setlength{\textfloatsep}{12pt plus 2pt minus 2pt}

We present the implementation details of the \FuncName{Combine}() function in Algorithm~\ref{algo:Combine} as below.

\noindent\textbf{Combine Wedge Sets}. \rev{Taking inspiration from the renowned Mergesort method~\cite{KatajainenPT96}, we adopt a recursive merging approach to combine the wedge sets in the wedge priority-increasing order. This method ensures that wedge permutations are efficient and balanced, and that only wedge pairs with different middle-vertices are checked.} As illustrated in Algorithm~\ref{algo:Combine}, given the hashmap $H[w]$ and the threshold $\delta$, \FuncName{Recur}() recursively merges two sets in a bottom-up manner until one set is left (line 1-7). Smart simultaneous implementations about set merging  (line 8-28) will be elaborated on in later paragraphs. \FuncName{Merge}() applies wedge priority as the merge rules and follows the original Mergesort method (line 29). 

\noindent\textbf{Order of Counting}. \rev{To ensure that the processing order of wedges in the merge process always follows the wedge priority,} for each subset, a pointer $ptr$ tracks the next wedge to process and a hashmap $HP$ maintains the visited wedges (line 9-11). Note that $HP$ only builds an array to store $t_a$ for each $t_s$. Subsequently, \AlgoName{TBC$^+$} identifies the maximum $t_s$ among all the unprocessed wedges, denoted as $maxn$ (lines 13-15). After filtering out illegal wedges, \AlgoName{TBC$^+$} query each unprocessed wedge with the maximum $t_s$ and the previous wedges in the $HP$ (line 19-25). Function \FuncName{Insert}() (line 26-28) updates the newly visited wedges in $HP$. Note that in actual implementation, \FuncName{Merge}() (line 29) can conduct in sync with \FuncName{Insert}() (line 26-28), as the \FuncName{Insert}() does not change $A$ and $D$. Specifically, \AlgoName{TBC$^+$} always processes wedges with the same $t_s$ together to avoid redundancy.

\setlength{\textfloatsep}{0pt}
\begin{algorithm}[tbp]
	\caption{\FuncName{Combine}() for Algorithm~\ref{algo:TBC+}}
	\label{algo:Combine}
	\linespread{0.8}\selectfont
	\LinesNumbered
	\DontPrintSemicolon
	\SetKwProg{Fn}{Function}{}{}
	\SetKwFunction{Recur}{\FuncName{Recur}}
	\SetKwFunction{SetCross}{\FuncName{SetCross}}
	\SetKwFunction{Insert}{\FuncName{Insert}}
	\SetKwFunction{Delete}{\FuncName{Delete}}
	\SetKwFunction{Enum}{\FuncName{Enum}}
	\SetKwFunction{Query}{\FuncName{Query}}
	\SetKwFunction{Merge}{\FuncName{Merge}}
	\KwIn{the start-vertex $u$; the threshold $\delta$; the hashmap $H[w]$ that including wedge sets $S_0, S_1, ..., S_{|H[w]|}$; the counts $\{C[i]\}_{i=0}^5$}
	\rev{\Recur{$u,\delta,0, |H[w]|, H[w], C[\cdot]$}\tcp*[r]{$C[\cdot]$ is $\{C[i]\}_{i=0}^5$}}
	\Fn{\rev{\Recur{$u,\delta,p, q, H[w], C[\cdot]$}}}{
		\textbf{if} $p + 1 \ge q$ \textbf{then} \textbf{return}\;
		$mid = (p + q) / 2$\;
		\rev{\Recur{$u,\delta,p, mid, H[w], C[\cdot]$}}\;
		\rev{\Recur{$u,\delta,mid, q, H[w], C[\cdot]$}}\;
		\rev{\SetCross{$u,\delta,S_p, S_{mid}, C[\cdot]$}}\;
	}
	\Fn{\rev{\SetCross{$u,\delta,S_i(A_i, D_i), S_j(A_j, D_j), C[\cdot]$}}}{
		\ForEach{$b \in \{A_i, D_i, A_j, D_j\}$}{
			$ptr_b := 0$\;
			initialize $HP_b$\;
		}
		\While{$\exists ptr_b < |b| : b \in \{A_i, D_i, A_j, D_j\}$}{
			$maxn := -\inf$\;
			\ForEach{$b \in \{A_i, D_i, A_j, D_j\} : ptr_b < |b|$}{
				$maxn := \max(maxn, s[ptr_b].t_s)$\;
			}
			\ForEach{$b \in \{A_i, D_i, A_j, D_j\}$}{
				\Delete{$maxn + \delta, HP_b$}\;
				$pre\_ptr_b = ptr_b$\;
			}
			\ForEach{$b \in \{A_i, D_i, A_j, D_j\}$}{
				\While{$ptr_b < |b| \wedge b[ptr_b].t_s = maxn$}{
					\rev{\textbf{if} $A_i$ \textbf{do} \Query{$u, b[ptr_b], HP_{A_j}, HP_{D_j}, C[\cdot]$}}\;
					\rev{\textbf{if} $D_i$ \textbf{do} \Query{$u, b[ptr_b], HP_{D_j}, HP_{A_j}, C[\cdot]$}}\;
					\rev{\textbf{if} $A_j$ \textbf{do} \Query{$u, b[ptr_b], HP_{A_i}, HP_{D_i}, C[\cdot]$}}\;
					\rev{\textbf{if} $D_j$ \textbf{do} \Query{$u, b[ptr_b], HP_{D_i}, HP_{A_i}, C[\cdot]$}}\;
					$ptr_b$ += $1$\;
				}
			}
			\ForEach{$b \in \{A_i, D_i, A_j, D_j\}$}{
				\For{$k := pre\_ptr_b$ \KwTo $ptr_b$}{
					\Insert{$s[k], HP_b$}\;
				}
			}
		}
		$S_i := ($\Merge{$A_i, A_j$}, \Merge{$D_i, D_j$}$)$\;
	}
\end{algorithm}
\setlength{\textfloatsep}{12pt plus 2pt minus 2pt}

\noindent\textbf{Counting Procedure}. \rev{To expedite the combination process between a wedge $\angle_i$ and multiple wedges ${\angle_j}$ satisfying $\angle_j.t_s > \angle_i.t_s$, we utilize the direct derivation $\angle.t_s < \angle.t_a \leq \angle.t_s + \delta$ from Lemma~\ref{lem:pruning}.} The set of wedges ${\angle_j}$ is maintained using a hashmap $HP$, which can be as simple as an array to store $t_a$ values for each $t_s$ encountered. Wedges that violate the duration constraint, as stated in Lemma~\ref{lem:bound}, i.e., $\angle_j$ with $\angle_j.t_a > \angle_i.t_s + \delta$, are eliminated. Lemma~\ref{lem:bound_d} guarantees that a deleted wedge from $HP$ will not be re-inserted. \rev{By appending wedges into $HP$ according to wedge priority, all the $t_a$ values in the same array in $HP$ are in ascending order. This enables us to utilize binary search to expedite the counting process.} In particular, $\angle_i.t_a<\angle_j.t_s<\angle_j.t_a$, $\angle_j.t_s < \angle_i.t_a < \angle_j.t_a$ and $\angle_j.t_s<\angle_j.t_a<\angle_i.t_a$ corresponding to case $c_{11}, c_{13}, c_{15}$ in Figure~\ref{fig:tbf_case}, respectively. Moreover, according to Lemma~\ref{lem:tbf_case11}, all the $t_a$ in $HP[t_s] : t_s > \angle_i.t_a$ has satisfied the case $c_{11}$ without binary search.

\begin{lemma}\label{lem:bound}
	Given the threshold $\delta$ and two wedges $\angle_i, \angle_j$ both satisfy $\angle.t_s < \angle.t_a \leq \angle.t_s + \delta$ and $\angle_i.t_s < \angle_j.t_s$, $\angle_i, \angle_j$ can form a temporal butterfly only if the condition $\angle_j.t_a \leq \angle_i.t_s + \delta$ is satified.
\end{lemma}

\FullVersion{\rev{\textsc{Proof}. Given the temporal duration constraint and $\angle_i.t_s$ is the minimum timestamp, this can be proven trivially. \hfill $\square$}}

\begin{lemma}\label{lem:bound_d}
	Given the threshold $\delta$ and three wedges $\angle_i, \angle_j, \angle_k$ all satisfy $\angle.t_s < \angle.t_a \leq \angle.t_s + \delta$ and $\angle_i.t_s < \angle_j.t_s < \angle_k.t_s$, if $\angle_k$ and $\angle_j$ can't form a temporal butterfly, then neither can $\angle_k$ and $\angle_i$.
\end{lemma}

\FullVersion{\rev{\textsc{Proof}. This can be proven immediately from the Lemma~\ref{lem:bound}. \hfill $\square$}}

\begin{lemma}\label{lem:tbf_case11}
	If two forward wedges $\angle_i, \angle_j$, $\angle_i.t_s < \angle_j.t_s$ satisfy $\angle_i.t_a<\angle_j.t_s$, we have $\angle_i.t_s < \angle_i.t_a < \angle_j.t_s <\angle_j.t_a$.
\end{lemma}

\FullVersion{\rev{\textsc{Proof}. This lemma is immediate. \hfill $\square$}}

Table~\ref{tab:api-hp} presents the operations supported by a hashmap $HP$ that maintains an ordered array for each key. Note that $HP[t].$\FuncName{pop}$(> x)$ run in the $O(n)$ time where $n$ denotes the number of elements to be popped out, and $HP[t].$\FuncName{count}$(\odot x)$ runs in $O(log|HP[t]|)$ time. All other operations consume $O(1)$ time.

\begin{table}[htbp]
	\vspace*{-0.1in}
	\centering
	\setlength{\tabcolsep}{2pt}
	\caption{Hashmap $HP$'s operations.}
	\vspace{-0.1in}
	\begin{tabular}{p{2.8cm}p{5cm}}
		\toprule[0.8pt]
		\textbf{API} & \textbf{Description} \\ \midrule
		$HP.$\FuncName{erase}$(t)$ & erase the $HP[t]$ \\ \midrule
		$|HP[t]|$ & return the size of $HP[t]$\\ \midrule
		\multirow{2}{*}{$HP[t].$\FuncName{empty}$()$} & return $true$ if $HP[t]$ is empty, \\
		& return $false$ otherwise\\ \midrule
		$HP[t].$\FuncName{append}$(x)$ & push $x$ into the back of $HP[t]$\\ \midrule
		$HP[t].$\FuncName{pop}$(> x)$ & pop all elements $> x$ \\ \midrule
		\multirow{2}{*}{$HP[t].$\FuncName{count}$(\odot x)$} & return the number of elements $\odot x$, \\
		& $\odot$ can be $<, >, \le, \ge$ \\
		\bottomrule[0.8pt]
	\end{tabular}\label{tab:api-hp}
	\vspace*{-0.1in}
\end{table}

Algorithm~\ref{algo:3func} illustrates the core functions in \FuncName{SetCross}(). Given the $bound$ (i.e., $maxn$+$\delta$ as above), \FuncName{Delete}() deletes all elements greater than $bound$ in the target hashmap and erases empty arrays (line 1-5). 
\FuncName{Query}() conducts a binary search to count all types of butterflies induced by a start-vertex and one of its wedges. 
The variable $l$ denotes the layer of $u$, and the type of butterflies can be converted through a simple xor operation $\oplus$ following our conversion rule (line 6-19). 
\FuncName{Insert}() simply append $\angle.t_a$ into the back of $HP[\angle.t_s]$, and this operation won't break the order in $HP$ (line 20-21).

\setlength{\textfloatsep}{0pt}
\begin{algorithm}[t]
	\caption{Core functions for Algorithm~\ref{algo:Combine}}
	\label{algo:3func}
	\linespread{0.8}\selectfont
	\LinesNumbered
	\DontPrintSemicolon
	\SetKwProg{Fn}{Function}{}{}
	\SetKwFunction{Insert}{\FuncName{Insert}}
	\SetKwFunction{Delete}{\FuncName{Delete}}
	\SetKwFunction{Query}{\FuncName{Query}}
	\SetKwFunction{Enum}{\FuncName{Enum}}
	\KwIn{the number $bound$; the hashmap $HP, HP_i, HP_j$; the start-vertex $u$; the wedge $\angle$; the counts $\{C[i]\}_{i=0}^5$}
	\Fn{\Delete{$bound, HP$}}{
		\ForEach{$t \in HP$}{
			$HP[t].$\FuncName{pop}$(> bound)$\;
			\If{$HP[t].$empty$()$}{$HP.$erase$(t)$\;}
		}
	}
	\Fn{\Query{$u, \angle, HP_i, HP_j, \{C[i]\}_{i=0}^5$}}{
		\textbf{if} $u \in U$ \textbf{then} $l := 0$ \textbf{else} $l := 1$\;
		\ForEach{$t \in HP_i$}{
			\uIf{$t > \angle.t_a$}{
				$C[0\oplus l]$ += $|HP_i[t]|$\tcp*[r]{$\oplus$ is xor operation}
			}
			\ElseIf{$t < \angle.t_a$}{
				$C[1\oplus l]$ += $HP_i[t].$\FuncName{count}$(> \angle.t_a)$\;
				$C[2\oplus l]$ += $HP_i[t].$\FuncName{count}$(< \angle.t_a)$\;
			}
		}
		\ForEach{$t \in HP_j$}{
			\uIf{$t > \angle.t_a$}{
				$C[3\oplus l]$ += $|HP_j[t]|$\;
			}
			\ElseIf{$t_s < \angle.t_a$}{
				$C[4\oplus l]$ += $HP_j[t].$\FuncName{count}$(> \angle.t_a)$\;
				$C[5\oplus l]$ += $HP_j[t].$\FuncName{count}$(< \angle.t_a)$\;
			}
		}
	}
	\Fn{\Insert{$\angle, HP$}}{
		$HP[\angle.t_s].$\FuncName{append}$(\angle.t_a)$\;
	}
	
\end{algorithm}
\setlength{\textfloatsep}{12pt plus 2pt minus 2pt}

\begin{figure}[htbp]
	\vspace*{-0.12in}
	\centering
	\includegraphics[width=3.0in]{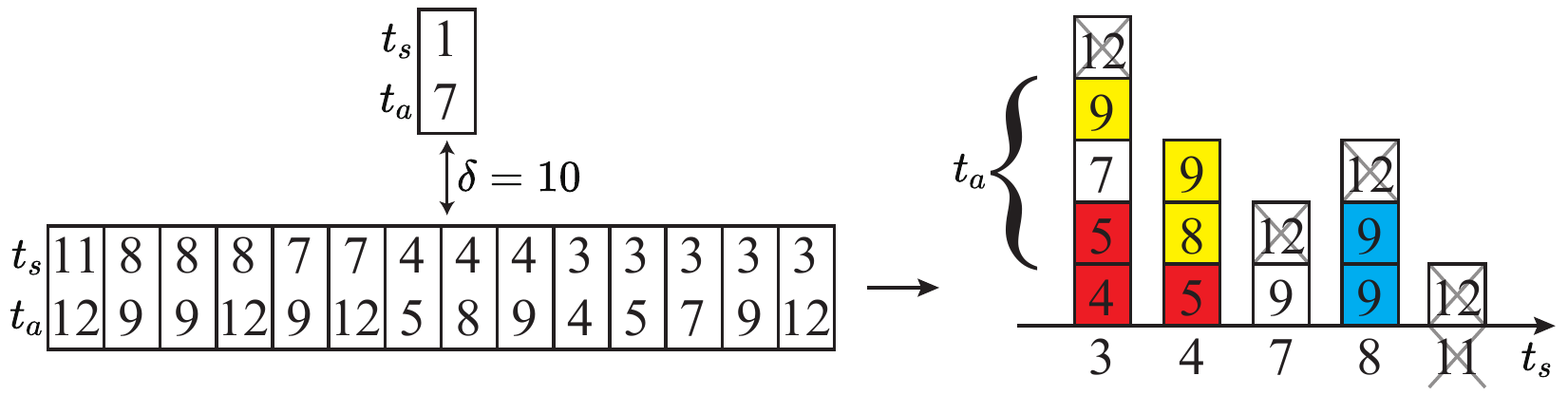}
	\vspace{-0.1in}
	\caption{Example for $HP$.}
	\label{fig:hp_case}
	\vspace*{-0.11in}
\end{figure}

\begin{example}
	We present a small example to illustrate $HP$ in Figure~\ref{fig:hp_case}. Suppose $\delta=10$, a wedge $\angle_i(1,7)$ and multiple wedges $\{\angle_j\} : \angle_j.t_s > \angle_i.t_s$ are sorted according to wedge priority. After inserting all wedges $\angle_j$ into $HP$, numbers under the axis denote $t_s$, and multiple squares on the axis denote the corresponding $t_a$. Then, squares with grey X represent wedges that need to be deleted, blue, yellow, and red these three kinds of squares respectively represent case $c_{11}, c_{13}, c_{15}$ after pairing with $\angle_i$.
\end{example}

\noindent\textbf{Complexity Analysis}.

\noindent$\bullet$ \rev{The time complexity of \AlgoName{TBC$^+$} is $O(\sum_{u\in V}|W(u)|(log(|E(u)|)+\alpha$\\$log(\frac{|W(u)|}{\alpha})))$, where $\alpha$ is a coefficient between 1 and $|W(u)|$.}

\textsc{Proof}. The main differences between \AlgoName{TBC} and \AlgoName{TBC$^+$} are in the counting phase. \rev{\FuncName{Recur}() involves each wedge asymptotic participate $log(|E(u)|)$ times in the \FuncName{SetCross}(), where $E(u)$ is the number of wedge sets. Within \FuncName{SetCross}(), wedges are traversed once, questioned once, inserted into $HP$ once, and deleted from $HP$ at most once. The time complexity of \FuncName{Insert}() and \FuncName{Delete}() are $O(1)$ but the \FuncName{Query}() is $O(\alpha log(\frac{n}{\alpha}))$, where $n$ denotes the number of wedges in $HP$, and $\alpha$ is a coefficient between 1 and $n$ depends on how many binary searches we run in \FuncName{Query}(). Due to the size of $HP$ under consideration, it's essential to evaluate the complexity of \FuncName{Query}() from a global perspective. In other words, for each wedge, distinct wedges with differing middle-vertices are considered precisely once in a \FuncName{Query}().} Thus, the total time complexity is proved. \hfill $\square$

\noindent$\bullet$ The space complexity of \AlgoName{TBC$^+$} is $O(|E|+\FuncName{max}_{u\in V}\{|W(u)|\}$.

\textsc{Proof}. The space complexity of the wedge enumerating process in \AlgoName{TBC$^+$} is essentially unchanged. While in the counting process, additional temporary auxiliary space is required for tracking merging sets and maintaining the $HP$. However, this never exceeds the original number of wedges. Thus, the space complexity remains $O(|E|+\FuncName{max}_{u\in V}\{|W(u)|\}$. \hfill $\square$

\subsection{Supporting Enumeration Algorithm}
\label{sec:enum}

In this section, we discuss the modifications made to our framework to enable butterfly enumeration. We introduce the inclusion of middle-vertex information in the wedge sets and the $HP$ hashmap, which was previously omitted during the counting process. This addition allows us to combine two wedges and obtain butterfly instances by determining the start- and end-vertices in advance. The overall procedure of \AlgoName{TBE$^+$} closely resembles that of \AlgoName{TBC$^+$}, with a variation in the \FuncName{Query}() function, as depicted in Algorithm~\ref{algo:enumerate}. In $HP$, wedges with the same $t_s$ are ordered based on their $t_a$ values. This ordering enables us to utilize range traversal to easily find specific types of temporal butterfly instances, similar to the binary search employed during the counting process. Specifically, in \AlgoName{TBE$^+$}, we iterate through $B[1\oplus l]$ from the beginning to the end, stopping as soon as the constraint is no longer satisfied (line 8-9). Similarly, we iterate through $B[2\oplus l]$ from the end to the beginning, breaking the loop once the constraint is violated (line 10-11). To enumerate all instances, the combination process is still required in \AlgoName{TBE$^+$}, which means the time and space complexity remains the same as \AlgoName{TBE}. However, due to efficient pruning strategies, we can eliminate the need for additional checks during the wedge combination, as discussed in \S~\ref{sec:baseline}. This allows us to directly determine the butterfly type, resulting in a significant improvement in efficiency, as demonstrated in \S~\ref{sec:exp}.

\setlength{\textfloatsep}{0pt}
\begin{algorithm}[tbp]
	\caption{\FuncName{Query}() for \AlgoName{TBE$^+$}}
	\label{algo:enumerate}
	\linespread{0.8}\selectfont
	\LinesNumbered
	\DontPrintSemicolon
	\SetKwProg{Fn}{Function}{}{}
	\SetKwFunction{Query}{\FuncName{Query}}
	\KwIn{the hashmap $HP_i, HP_j$; the start-vertex $u$; the wedge $\angle$; the butterfly instances $\{B[i]\}_{i=0}^5$}
	\Fn{\Query{$u, \angle, HP_i, HP_j, \{B[i]\}_{i=0}^5$}}{
		\textbf{if} $u \in U$ \textbf{then} $l := 0$ \textbf{else} $l := 1$\;
		\ForEach{$t \in HP_i$}{
			\uIf{$t > \angle.t_a$}{
				\ForEach{$\angle_j \in HP_i[t]$}{$B[0\oplus l]$.\FuncName{append}$((\angle, \angle_j))$\;}
			}
			\ElseIf{$t < \angle.t_a$}{
				\ForEach{$\angle_j \in HP_i[t]:\angle_j.t_a>\angle.t_a$}{$B[1\oplus l]$.\FuncName{append}$((\angle, \angle_j))$\;}
				\ForEach{$\angle_j \in HP_i[t]:\angle_j.t_a<\angle.t_a$}{$B[2\oplus l]$.\FuncName{append}$((\angle, \angle_j))$\;}
			}
		}
		\ForEach{$t \in HP_j$}{
			\tcp*[l]{handle $B[3/4/5\oplus l]$ similar to line 4-11}
		}
	}
\end{algorithm}
\setlength{\textfloatsep}{12pt plus 2pt minus 2pt}

\vspace{-0.05in}
\subsection{Handling Extreme Cases}
\label{sec:opt+}

\begin{figure}[htbp]
	\vspace*{-0.1in}
	\centering
	\includegraphics[width=1.6in]{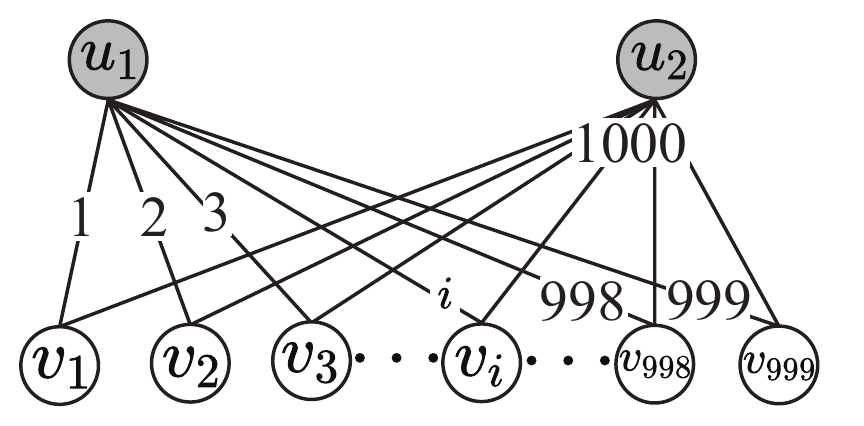}
	\vspace{-0.1in}
	\caption{A temporal bipartite graph containing two high-degrees vertices $u_1$ and $u_2$.}
	\label{fig:sp_case}
	\vspace*{-0.1in}
\end{figure}

\rev{Notably, the core bottleneck of \AlgoName{TBC$^+$} lies in the unstable efficiency of \AlgoName{Query}(). Figure~\ref{fig:sp_case} presents an extreme case while $u_1$ is the start-vertex and all the wedges have different $t_s$, leading to a quadratic time in wedge combinations (i.e., $\alpha \approx |W(u)|$ in the time complexity of \AlgoName{TBC$^+$}).} This is a common situation in real-world datasets: a small number of vertices come with a very high degree (and subsequently many wedges with different $t_s$)~\cite{clauset2009power, onnela2007structure}.

We further equip our counting solution with two red-black trees~\cite{hinze1999constructing, cormen2022introduction} to resolve the issue. Specifically, $TA$ is a red-black tree to maintain wedges with the key $t_a$, and $TS$ is a twin red-black tree of $TA$ that only maintains $t_s$ with the key $t_s$. These two trees are synchronized and contain the same elements, but they are organized based on different keys. This design allows us to perform efficient two-way operations compared to storing all the elements in a single hashmap. All operations of $TA, TS$ (as shown in Table~\ref{tab:rbtree}) run in $O(log(n))$ time, where $n$ is the number of elements in a tree.

\begin{table}[htbp]
	\centering
	\setlength{\tabcolsep}{2pt}
	\caption{Red-black tree $TA/TS$'s operations.}
	\vspace{-0.1in}
	\begin{tabular}{p{2.6cm} p{5.4cm}}
		\toprule[0.8pt]
		\textbf{API} & \textbf{Description} \\ \midrule
		$TA/TS.$\FuncName{back}$()$ & return the last $\angle/t_s$ in $TA/TS$\\ \midrule
		\multirow{2}{*}{$TA/TS.$\FuncName{count}$(\odot x)$} & return the number of elements with \\
		& the key $\odot x$, $\odot$ can be $<, >, \le, \ge$ \\ \midrule
		$TA/TS.$\FuncName{erase}$(\angle/t_s)$ & erase a $\angle/t_s$ from $TA/TS$ \\ \midrule
		$TA/TS.$\FuncName{insert}$(\angle/t_s)$ & insert a $\angle/t_s$ into $TA/TS$ \\
		\bottomrule[0.8pt]
	\end{tabular}\label{tab:rbtree}
\end{table}

The core differences lie in speeding up the \FuncName{Query}() and \FuncName{Delete}() functions with two red-black trees $TA, TS$ (in \AlgoName{TBC$^{++}$}) instead of using hashmap $HP$. We omit minor modifications on \FuncName{SetCross}() due to space limitations. Algorithm~\ref{algo:3func+} demonstrates the improvements made to Algorithm~\ref{algo:3func}. When deleting wedges, following Lemma~\ref{lem:bound}, \AlgoName{TBC$^{++}$} checks the last elements in $TA$ and erases it from both $TA$ and $TS$ if the condition is violated (line 1-4). Based on Lemma~\ref{lem:tbf_case11}, \AlgoName{TBC$^{++}$} conducts \FuncName{count}() on $TS$ which corresponds to the case $c_{11}$ (line 7, 10). Similarly, according to Lemma~\ref{lem:tbf_case131} and Lemma~\ref{lem:tbf_case132}, \AlgoName{TBC$^{++}$} conduct \FuncName{count}() on both $TA, TS$ and subtract which corresponds the case $c_{13}$ (line 8, 11). According to Lemma~\ref{lem:tbf_case15}, \AlgoName{TBC$^{++}$} does \FuncName{count}() on $TA$ which corresponds the case $c_{15}$ (line 9, 12). 

\begin{lemma}\label{lem:tbf_case15}
	If two forward wedges $\angle_i, \angle_j$, $\angle_i.t_s < \angle_j.t_s$ satisfy $\angle_j.t_a<\angle_i.t_a$, we have $\angle_i.t_s < \angle_j.t_s < \angle_j.t_a <\angle_i.t_a$.
\end{lemma}

\FullVersion{\rev{\textsc{Proof}. This lemma is immediate. \hfill $\square$}}

\begin{lemma}\label{lem:tbf_case131}
	If two forward wedges $\angle_i, \angle_j$, $\angle_i.t_s < \angle_j.t_s$ satisfy $\angle_i.t_a < \angle_j.t_a$, we have $\angle_i.t_s < \angle_i.t_a \le \angle_j.t_s <\angle_j.t_a$ or $\angle_i.t_s < \angle_j.t_s <\angle_i.t_a<\angle_j.t_a$.
\end{lemma}

\FullVersion{\rev{\textsc{Proof}. This lemma is immediate. \hfill $\square$}}

\begin{lemma}\label{lem:tbf_case132}
	If two forward wedges $\angle_i, \angle_j$, $\angle_i.t_s < \angle_j.t_s$ satisfy $\angle_i.t_a \le \angle_j.t_s$, we have $\angle_i.t_s < \angle_i.t_a \le \angle_j.t_s <\angle_j.t_a$.
\end{lemma}

\FullVersion{\rev{\textsc{Proof}. This lemma is immediate. \hfill $\square$}}

\setlength{\textfloatsep}{0pt}
\begin{algorithm}[t]
	\caption{Core functions for \AlgoName{TBC$^{++}$}}
	\label{algo:3func+}
	\linespread{0.8}\selectfont
	\LinesNumbered
	\DontPrintSemicolon
	\SetKwProg{Fn}{Function}{}{}
	\SetKwFunction{Insert}{\FuncName{Insert}}
	\SetKwFunction{Delete}{\FuncName{Delete}}
	\SetKwFunction{Query}{\FuncName{Query}}
	\KwIn{the number $bound$; the red-black tree $TS, TS_i, TS_j$ with key $t_s$; the red-black tree $TA, TA_i, TA_j$ with key $t_a$; the start-vertex $u$; the wedge $\angle$; the counts $\{C[i]\}_{i=0}^5$}
	\Fn{\Delete{$bound, (TS, TA)$}}{
		\While{$\angle := TA.$\FuncName{back}$() : \angle.t_a > bound$}{
			$TA.$\FuncName{erase}$(\angle)$\;
			$TS.$\FuncName{erase}$(\angle.t_s)$\;
		}
	}
	\Fn{\Query{$u, \angle, (TS_i, TA_i), (TS_j, TA_j), \{C[i]\}_{i=0}^5$}}{
		\textbf{if} $u \in U$ \textbf{then} $l := 0$ \textbf{else} $l := 1$\;
		$C[0\oplus l]$ += $TS_i.$\FuncName{count}$(> \angle.t_a)$\;
		$C[1\oplus l]$ += $TA_i.$\FuncName{count}$(> \angle.t_a) - TS_i.$\FuncName{count}$(\ge \angle.t_a)$\;
		$C[2\oplus l]$ += $TA_i.$\FuncName{count}$(< \angle.t_a)$\;
		$C[4\oplus l]$ += $TS_j.$\FuncName{count}$(> \angle.t_a)$\;
		$C[5\oplus l]$ += $TA_j.$\FuncName{count}$(> \angle.t_a) - TS_j.$\FuncName{count}$(\ge \angle.t_a)$\;
		$C[6\oplus l]$ += $TA_j.$\FuncName{count}$(< \angle.t_a)$\;
	}
	\Fn{\Insert{$\angle, (TS, TA)$}}{
		$TS.$\FuncName{insert}$(\angle)$\;
		$TA.$\FuncName{insert}$(\angle.t_s)$\;
	} 
\end{algorithm}
\setlength{\textfloatsep}{12pt plus 2pt minus 2pt}

\begin{figure}[htbp]
	\vspace*{-0.1in}
	\centering
	\includegraphics[width=3.2in]{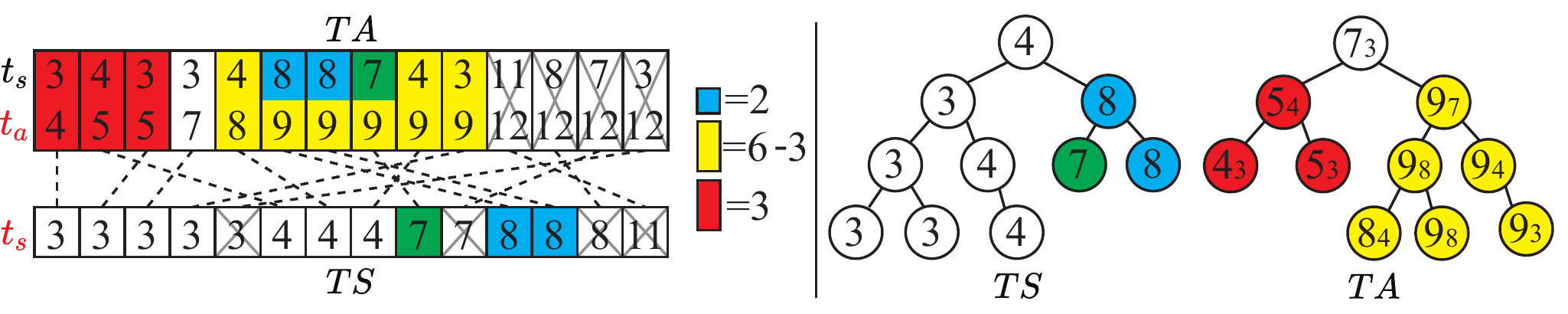}
	\vspace{-0.1in}
	\caption{Example for $TA, TS$.}
	\label{fig:tsta_case}
	\vspace*{-0.1in}
\end{figure}

\begin{example}
	In Figure~\ref{fig:tsta_case}, we show how to use $TA, TS$ to handle the query in Figure~\ref{fig:hp_case}. Suppose two ordered lists represent the red-black trees, keeping the same key order as in the corresponding tree. The elements in the two trees align with each other via dotted lines. Then rectangles with grey X represent wedges that need to be deleted, and corresponding squares with grey X will be deleted as well. Blue squares, pure yellow rectangles, and red rectangles are these three kinds of polygons representing case $c_{11}$, $c_{13}$, and $c_{15}$, respectively.
\end{example}

\noindent\textbf{Complexity Analysis}. 

\noindent$\bullet$ \rev{The time complexity of \AlgoName{TBC$^{++}$} is $O(\sum_{u\in V}|W(u)|(log(|E(u)|)+log(|W(u)|)))$.}

\textsc{Proof}. Note that, \FuncName{Delete}(), \FuncName{Insert}() and \FuncName{Query}() are all run in $O(log(n))$, where n is the number of wedges in $TA$/$TS$. Thus, the time complexity of \AlgoName{TBC$^{++}$} is proven. \hfill $\square$

\noindent$\bullet$ The space complexity of \AlgoName{TBC$^{++}$} is $O(|E|+\FuncName{max}_{u\in V}\{|W(u)|\}$.

\textsc{Proof}. Although $TS$ and $TA$ require constant times the memory cost of $HP$, the space complexity remains unchanged. \hfill $\square$

%% file: 5-Streaming.tex
\section{Algorithms on Graph Streams}
\label{sec:stream}

\begin{figure}[htbp]
	\vspace*{-0.1in}
	\centering
	\includegraphics[width=2.8in]{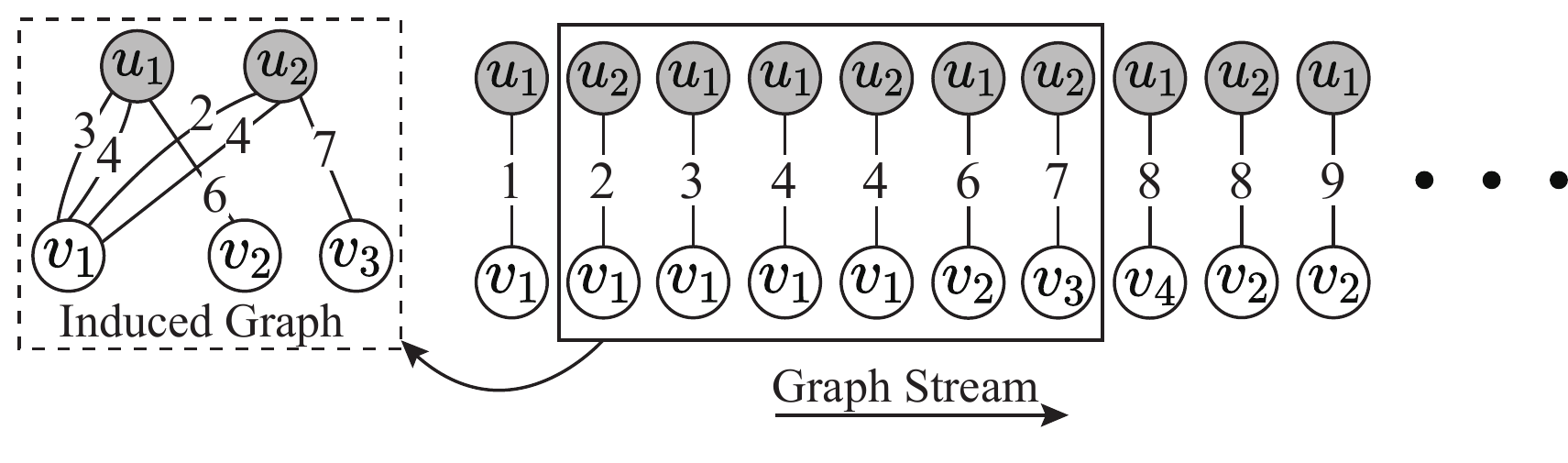}
	\vspace{-0.1in}
	\caption{A temporal bipartite graph stream.}
	\label{fig:streaming}
	\vspace*{-0.1in}
\end{figure}

Temporal bipartite graphs often encounter frequent rapid updates, involving the insertion or deletion of vertices and edges. These updates are commonly represented as graph streams, where edges are presented in their timestamp order \cite{mcgregor2014graph}. To efficiently count temporal butterflies, it is preferable to update the counts incrementally instead of recomputing them from scratch. Figure~\ref{fig:streaming} illustrates a temporal bipartite graph stream, with the corresponding induced graph of selected edges displayed on the left-hand side. 
While it is possible to extend many of the mentioned algorithms to graph streams, we will focus our discussion on the most effective counting algorithm (i.e., \AlgoName{TBC$^{++}$}), for the sake of brevity. 

\rev{The intuitive idea is to count the number of temporal butterflies that contain the edge waiting update and modify the counts accordingly. In this context, we propose the \AlgoName{STBC} algorithm \PubVersion{(detailed in~\cite{cai2023efficient})}\FullVersion{(detailed in Algorithm~\ref{algo:stbc})}, a non-trivial extension from \AlgoName{TBC$^{++}$}. The major changes are as follows: \textbf{(1)} While enumerating wedges, vertex priority becomes irrelevant as all butterflies induced by the current edge must be considered, but Lemma~\ref{lem:pruning} still applies as a timestamp $t$ is known. If we store $E(u)$ in a queue and process it in chronological order, we can use binary search to compress the traversal range into $[t$-$\delta, t$+$\delta]$.
\textbf{(2)} Since a middle-vertex $v$ is determined, to avoid unnecessary merging operations, we only need to maintain two wedge sets within each $H[w]$ depending on whether the middle-vertex is $v$ or not.}

\FullVersion{
\setlength{\textfloatsep}{0pt}
\begin{algorithm}[tbp]
    \caption{\AlgoName{STBC} (delete a single edge) }
    \label{algo:stbc}
    \linespread{0.8}\selectfont
    \LinesNumbered
    \DontPrintSemicolon
    \SetKwProg{Fn}{Function}{}{}
    \SetKwFunction{SetCross}{\FuncName{SetCross}}
    \KwIn{the temporal bipartite graph $G=(V = (U, L), E)$; the threshold $\delta$; the edge waiting for delete $e(u, v, t)$; the counts $\{C[i]\}_{i=0}^5$}
    initialize hashmap $H$ for each $w$ to store sets\;
    \ForEach{$(u, x, t') \in E(u) : t-\delta\le t'\le t+\delta$}{
        \If{$x\neq v \wedge t'\neq t$}{
            \ForEach{$(x, w, t'') \in E(x)$ : $\FuncName{max}\{t,t'\}-\delta\le t''\le \FuncName{min}\{t,t'\}+\delta$}{
                \If{$w\neq u \wedge t''\neq t \wedge t''\neq t'$}{
                    \uIf{$t' < t''$}{
                        $H[w][!v].A.$\FuncName{append}$(\angle(t', t''))$\;
                    }
                    \ElseIf{$t' > t''$}{
                        $H[w][!v].D.$\FuncName{append}$(\angle(t'', t'))$\;
                    }
                }
            }
        }
        
    }
    \ForEach{$(v, w, t') \in E(v) : t-\delta \le t'\le t+\delta$}{
        \If{$w\neq u \wedge t'\neq t$}{
            \uIf{$t < t'$}{
                $H[w][v].A.$\FuncName{append}$(\angle(t, t'))$\;
            }
            \ElseIf{$t > t'$}{
                $H[w][v].D.$\FuncName{append}$(\angle(t', t))$\;
            }
        }
    }
    \ForEach{vertex $w \in H : |H[w]| > 1$}{
        sort each subsets in $H[w]$ according to $P_W(\angle)$\;
        $\{C[i]\}_{i=0}^5$ -= \SetCross{$u, \delta, H[w][!v], H[w][v]$}\;
    }
    delete $e$ from $G$\;
\end{algorithm}
\setlength{\textfloatsep}{12pt plus 2pt minus 2pt}
}

We also propose a parallel version of the algorithm, called \AlgoName{STBC$^+$}, which is presented in Algorithm~\ref{algo:stbc+}. \rev{Regarding edge deletion, in accordance with Lemma~\ref{lem:batch-update}, \AlgoName{STBC$^+$} redefines the traversal range to $(t,t$+$\delta]$, thereby preventing count conflicts. Noted that $t$ represents the minimum timestamp in a temporal butterfly, and the temporal duration can be easily satisfied while enumerating wedges. Consequently, \AlgoName{STBC$^+$} no longer needs red-black trees $TS, TA$ to maintain wedges dynamically, but only two simple arrays $VS, VA$. During combination, \AlgoName{STBC$^+$} only sorts these $VS, VA$ in need and performs \FuncName{Query}() in a similar manner as in \AlgoName{TBC$^{++}$}. To prevent read-write conflicts between threads, the deletion of edges is executed after the counting process is completed.} Similarly, when inserting edges, the traversal range is redefined as $[t$-$\delta,t)$, and all edges should be inserted into the graph beforehand.

\setlength{\textfloatsep}{0pt}
\begin{algorithm}[tbp]
	\caption{\AlgoName{STBC$^+$} (delete multiple edges)}
	\label{algo:stbc+}
	\linespread{0.8}\selectfont
	\LinesNumbered
	\DontPrintSemicolon
	\SetKwProg{Fn}{Function}{}{}
	\SetKwFunction{Query}{\FuncName{Query}}
	\KwIn{the temporal bipartite graph $G=(V = (U, L), E)$; the threshold $\delta$; the edges waiting for delete $\{e_1, e_2, \cdots, e_i\}$, the counts $\{C[i]\}_{i=0}^5$}
	\ForEach{$e(u, v, t) \in \{e_1, e_2, \cdots, e_i\}$ \textbf{in parallel}}{
		initialize hashmap $H$ for each $w$ to store sets\;
		\ForEach{$(u, x, t') \in E(u) : t<t'\le t+\delta$}{
			\If{$x\neq v \wedge t'\neq t$}{
				\ForEach{$(x, w, t'') \in E(x) : t<t''\le t+\delta$}{
					\If{$w\neq u$}{
						\uIf{$t' < t''$}{
							$H[w].A.VS.$\FuncName{append}$(t')$\;
							$H[w].A.VA.$\FuncName{append}$(t'')$\;
						}
						\ElseIf{$t' > t''$}{
							$H[w].D.VS.$\FuncName{append}$(t'')$\;
							$H[w].D.VA.$\FuncName{append}$(t')$\;
						}
					}
				}
			}
			
		}
		\ForEach{$(v, w, t') \in E(v) : t<t'\le t+\delta$}{
			\If{$w\neq u$}{
				\If{$H[w]$ is unsorted}{sort $A.VS, A.VA, D.VS, D.VA$ in $H[w]$\;}
				$\{C[i]\}_{i=0}^5$ -= \Query{$u,(t,t'), H[w].A, H[w].D$}\;
			}
		}
	}
	\ForEach{$e \in \{e_1, e_2, \cdots, e_i\}$}{
		delete $e$ from $G$\;
	}
\end{algorithm}
\setlength{\textfloatsep}{12pt plus 2pt minus 2pt}

\begin{lemma}\label{lem:batch-update}
	When counting contained in a batch of edges with the minimum/maximum time, counting each temporal butterfly on the edge with the minimum/maximum timestamp can prevent count conflicts.
\end{lemma}
\HidProf{\textsc{Proof}. A temporal butterfly only has one minimum/maximum timestamp, and thus can avoid being counted multiple times. \hfill $\square$}

\noindent\textbf{Complexity Analysis}.

\noindent$\bullet$ Given the edge $e(u,v,t)$, the time complexity of updating a single edge in \AlgoName{STBC} and \AlgoName{STBC$^+$} are both $O(|E^2(u)|log(|E^2(u)|))$, where $|E^2(u)|=\sum_{(u,v,t)\in E(u)}|E(v)|$.

\textsc{Proof}. The edges enumerated by \AlgoName{STBC} and \AlgoName{STBC$^+$} is denoted by $E^2(u)$, where $|E^2(u)|=\sum_{(u,v,t)\in E(u)}|E(v)|$, thus the time complexity are both $O(|E^2(u)|log(|E^2(u)|))$ similar to the complexity analysis of \AlgoName{TBC$^{++}$}. Despite \AlgoName{STBC} and \AlgoName{STBC$^+$} having the same time complexity, \AlgoName{STBC$^+$} has a smaller constant than \AlgoName{STBC} since \AlgoName{STBC$^+$} uses two simple arrays to replace the red-black tree. \hfill $\square$

\noindent$\bullet$ Given the edge $e(u,v,t)$, the time complexity of updating a single edge in \AlgoName{STBC} and \AlgoName{STBC$^+$} are both $O(|E|+|E^2(u)|$, where $|E^2(u)|=\sum_{(u,v,t)\in E(u)}|E(v)|$.

\textsc{Proof}. As noted in the proof for time complexity, the edges enumerated by \AlgoName{STBC} and \AlgoName{STBC$^+$} are denoted by $E^2(u)$, where $|E^2(u)|=\sum_{(u,v,t)\in E(u)}|E(v)|$. \hfill $\square$

%% file: 6-Exp.tex
\vspace{-0.03in}
\section{Experimental Evaluation}
\vspace{-0.01in}
\label{sec:exp}

\begin{table}[tbp]
    \vspace*{-0.1in}
    \centering
    \setlength{\tabcolsep}{2pt}
    \caption{The summary of datasets.}
    \vspace{-0.1in}
    \begin{tabular}{c||c|c|c|c}
        \hline
        \multirow{2}{*}{\textbf{Dataset}} & \multirow{2}{*}{$|E|$} & \multicolumn{2}{c|}{$|V|$} & \textbf{Time Span}\\ \cline{3-4}
        & & $|U|$ & $|L|$ & (days)\\ \hline \hline
        Wikiquote (WQ) & 776,458 & 961 & 640,482 & 4625.66 \\ \hline
        Wikinews (WN) & 907,499 & 2,200 & 35,979 & 4857.34 \\ \hline
        StackOverflow (SO) & 1,301,942 & 545,196 & 96,680 & 1153.00 \\ \hline
        CiteULike (CU) & 2,411,819 & 153,277 & 731,769 & 1203.10 \\ \hline
        Bibsonomy (BS) & 2,555,080 & 204,673 & 767,447 & 7665.43 \\ \hline
        Twitter (TW) & 4,664,605 & 175,214 & 530,418 & 1155.34\\ \hline
        Amazon (AM) & 5,838,041 & 2,146,057 & 1,230,915 & 3650.00 \\ \hline
        Edit-ru (ER) & 8,349,235 & 7,816 & 1,266,349 & 4976.35 \\ \hline
        Epinions (EP) & 13,668,320 & 120,492 & 755,760 & 504.96 \\ \hline
        Last.fm (LF) & 19,150,868 & 992 & 174,077 & 3149.77\\ \hline
        Wiktionary (WT) & 44,788,448 & 66,140 & 5,826,113 & 5941.22 \\ \hline
    \end{tabular}\label{tab:dataset}
    \vspace*{-0.13in}
\end{table}

In this section, we present the empirical evaluation of our solutions using 11 large-scale real-world datasets. 

\noindent\rev{\textbf{Experiment Settings}}. All our algorithms\footnote{Available at  https://github.com/ZJU-DAILY/TBFC} were implemented in C++ and executed on a Ubuntu machine with an Intel(R) Core(TM) i9-10900K CPU @ 3.70GHz and 128G memory. We set a maximum running time limit of 100,000 seconds and terminate the execution if the limit is exceeded. Notably, the reported time costs do not include preprocessing time, such as the graph loading time. The space cost is measured by monitoring the maximum VmRSS (Virtual Memory Resident Set Size) of the process.

\noindent\rev{\textbf{Algorithms}. The competitors include: 
\textbf{(1)} \underline{T}emporal \underline{B}utterfly \underline{C}oun-ting algorithms: \AlgoName{TBC}, \AlgoName{TBC$^+$}, and \AlgoName{TBC$^{++}$}. 
\textbf{(2)} \underline{T}emporal \underline{B}utterfly \underline{E}numeration algorithms: \AlgoName{TBE} and \AlgoName{TBE$^+$}. 
\textbf{(3)} \underline{S}treaming \underline{T}emporal \underline{B}utterfly \underline{C}ounting algorithms: \AlgoName{STBC} and \AlgoName{STBC$^+$}. 
\textbf{(4)} two temporal motif isomorphism algorithms~\cite{mackey2018chronological, li2019time} and a temporal motif counting algorithm~\cite{paranjape2017motifs}.
In the evaluation of enumeration algorithms, we do not perform any additional actions, such as outputting butterfly instances to external memory, when they are found. This is because directly storing instances in external memory would introduce additional time costs due to I/O operations while storing them in RAM would result in extra space costs. To ensure fair comparison experiments, we focus solely on the enumeration process. 
In addition to our proposed algorithms for temporal butterflies, we also attempted some state-of-the-art general temporal motif algorithms~\cite{mackey2018chronological, li2019time, paranjape2017motifs} in accordance with the same experimental setup. However, it is important to note that no algorithm specifically designed for temporal butterflies exists. Even on datasets considered ``easy to handle'', such as WQ, SO, and CU, our algorithm completed within 10 seconds (Figure~\ref{fig:exp1_time}), whereas the general algorithms failed to meet the time limit due to the need to permute all possible combinations of four edges in the worst case. Consequently, we have excluded this comparison from our evaluation.}

\noindent\rev{\textbf{Datasets}}. The dataset statistics are presented in Table~\ref{tab:dataset}, where "Time Span" indicates the time difference between the maximum and minimum timestamps. In our study of the temporal bipartite graph stream, we assume that edges arrive in chronological order. For the purpose of evaluation, we adopt the widely used Sliding Window Model~\cite{kim2022denforest} for \ProbName{streaming temporal butterfly counting}. This involves counting butterflies within a window of size $|window|$ while sliding with a stride of size $|stride|$ at each step. Both $|window|$ and $|stride|$ are measured in terms of the number of edges. Additional dataset sources and more detailed statistics can be found at KONECT\footnote{http://konect.cc/}.

\vspace{-0.1in}
\subsection{Evaluation on Temporal Bipartite Graphs}
\begin{figure*}[t]
    \vspace*{-0.1in}
    \begin{minipage}[t]{0.48\linewidth}
        \centering
        \includegraphics[width=\textwidth]{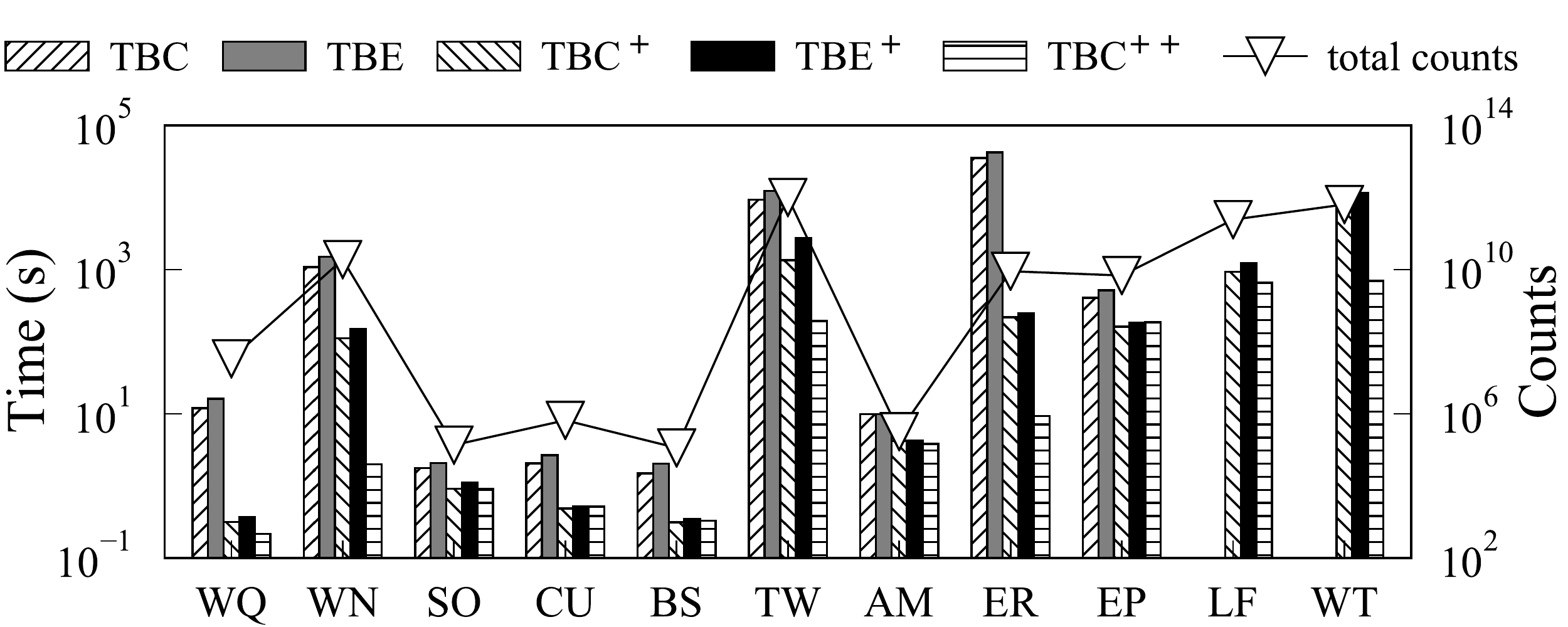}
        \vspace*{-0.25in}
        \caption{Time and total counts on varying datasets.}
        \label{fig:exp1_time}
    \end{minipage}\hspace*{0.05in}
    \begin{minipage}[t]{0.48\linewidth}
        \centering
        \includegraphics[width=\textwidth]{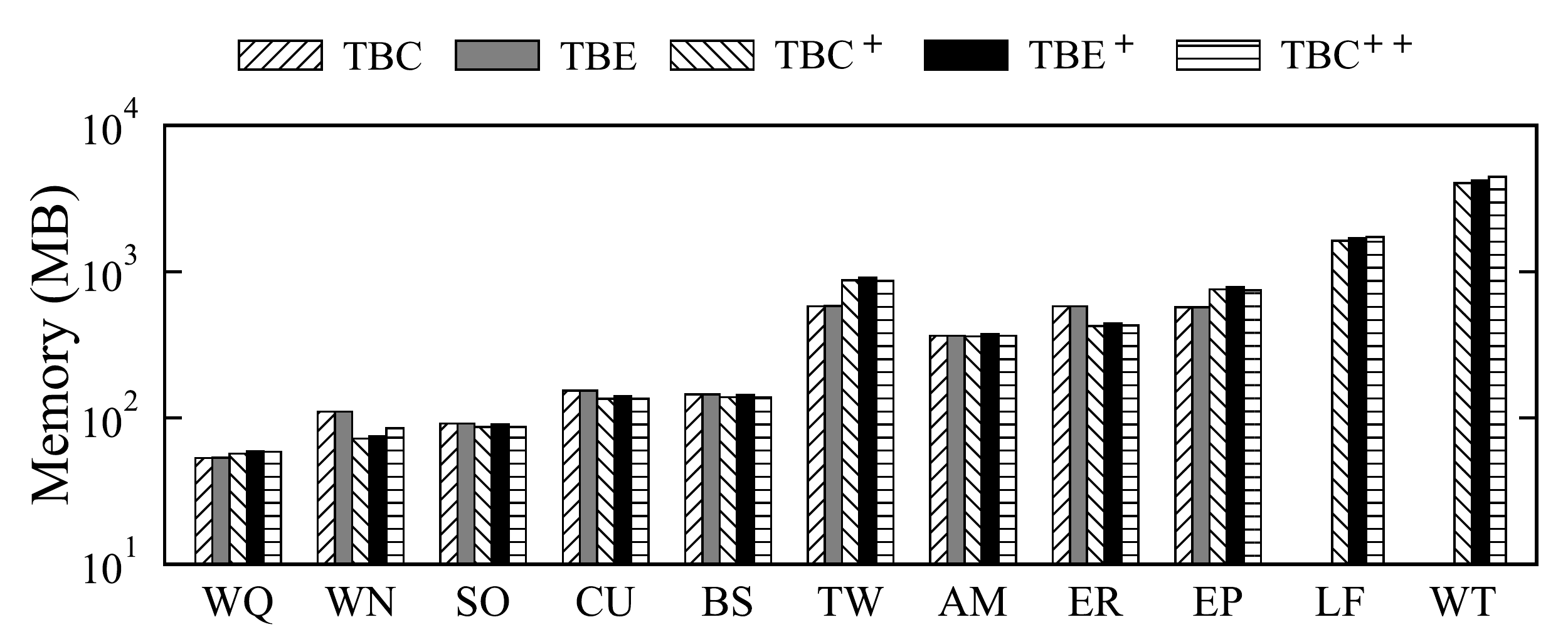}
        \vspace*{-0.25in}
        \caption{Memory on varying datasets.}
        \label{fig:exp1_mem}
    \end{minipage}
    \vspace*{-0.1in}
\end{figure*}

\vspace{-0.02in}
\noindent\textbf{Overall Performance}. The efficiency of our baseline algorithms \AlgoName{TBC} and \AlgoName{TBE}, as well as our three optimization versions \AlgoName{TBC$^+$}, \AlgoName{TBE$^+$}, and \AlgoName{TBC$^{++}$}, is compared on various datasets as shown in Figure~\ref{fig:exp1_time}, with a default $\delta$ value of 40 days.

\begin{figure*}[t]
    \centering
    \includegraphics[width=0.32\textwidth]{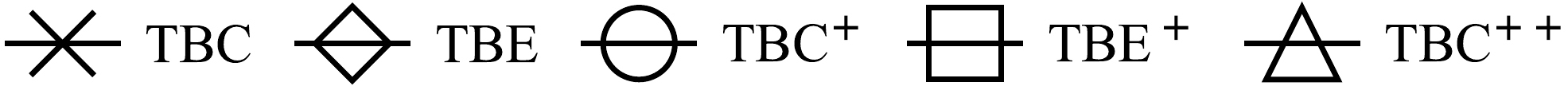}\\
    \vspace*{-0.09in}
    \subfigure[WN]{
        \label{fig:exp2_WN}
        \includegraphics[width=0.245\textwidth]{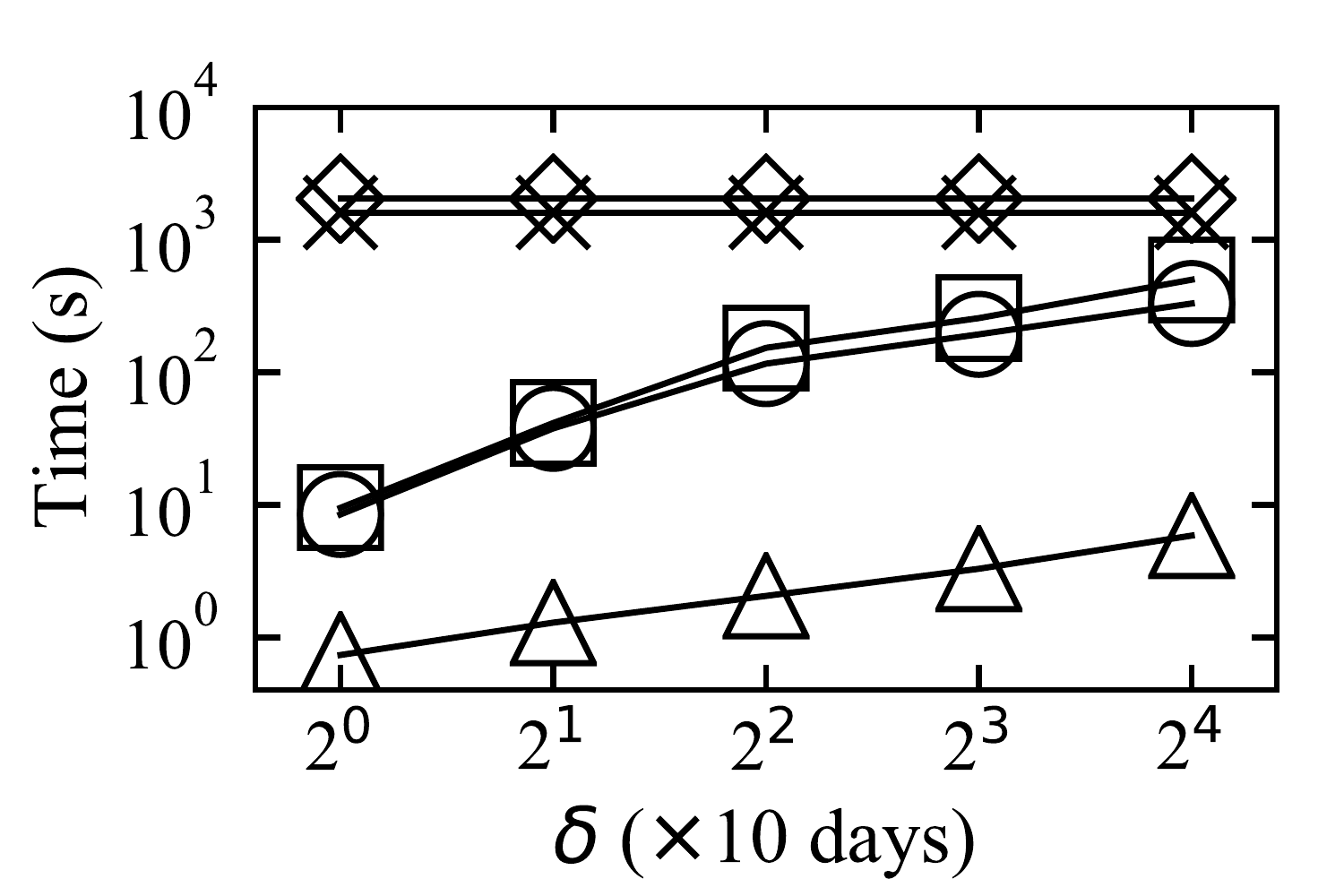}
    }\hspace{-3mm}
    \subfigure[AM]{
        \label{fig:exp2_AM}
        \includegraphics[width=0.245\textwidth]{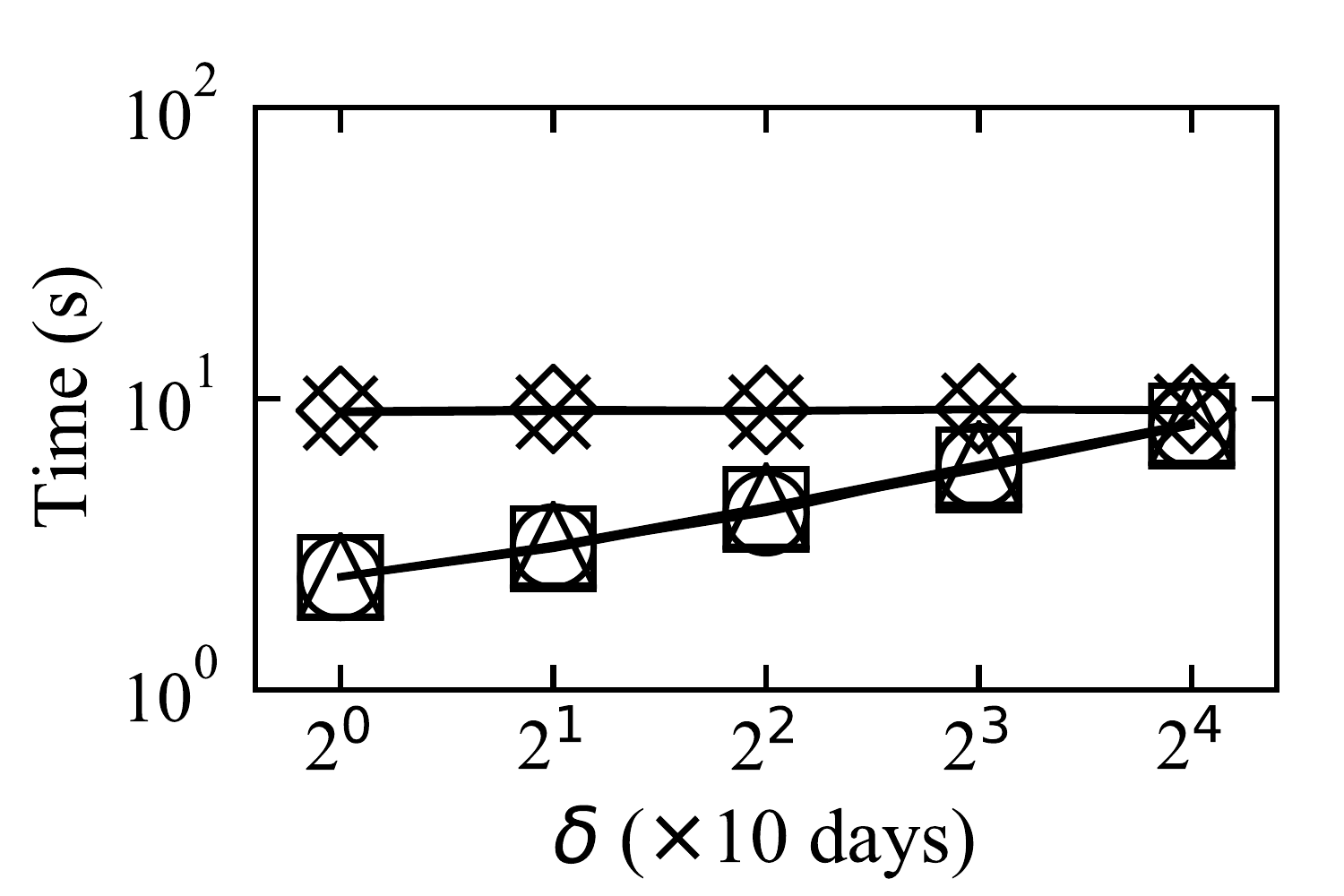}
    }\hspace{-3mm}
    \subfigure[ER]{
        \label{fig:exp2_ER}
        \includegraphics[width=0.245\textwidth]{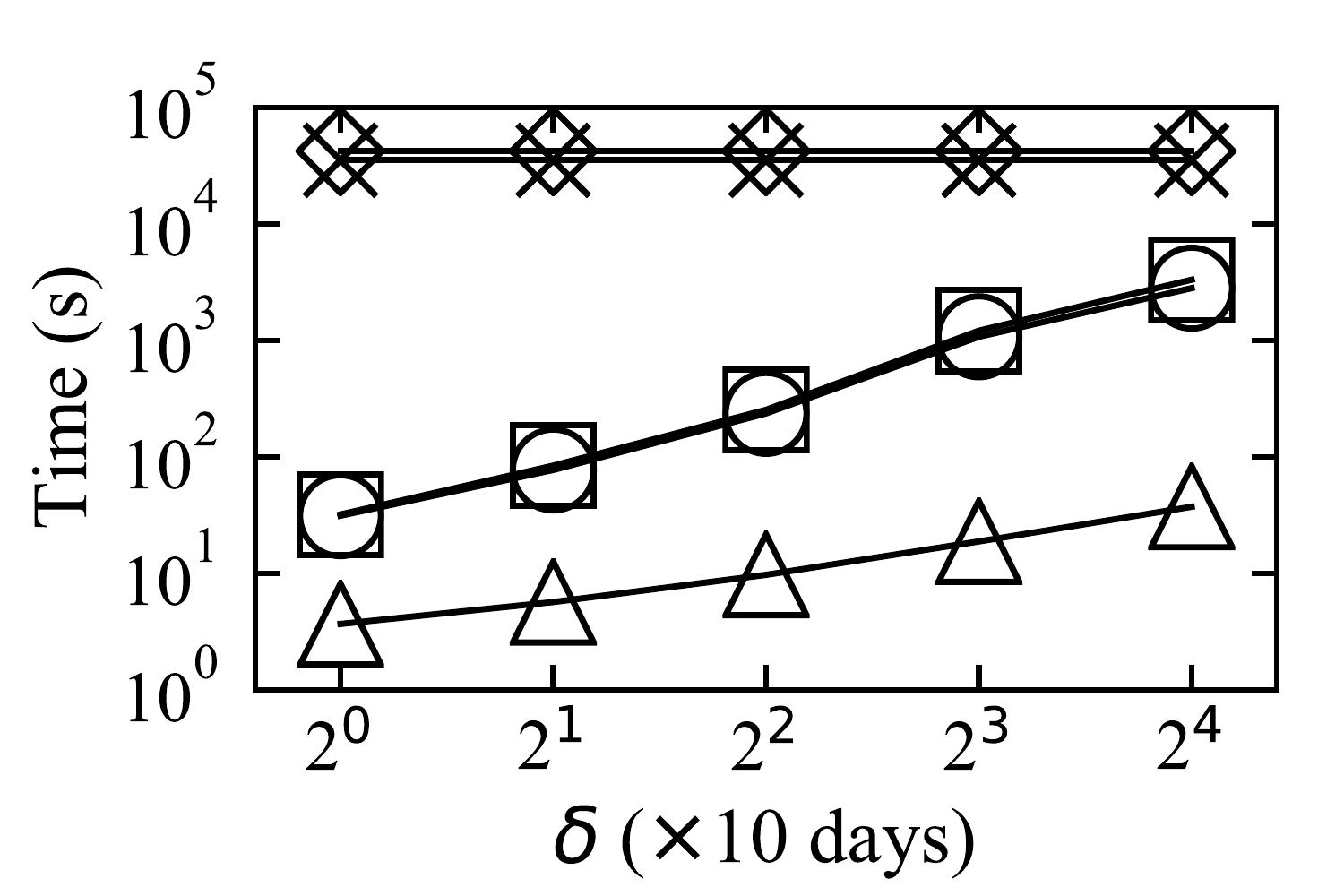}
    }\hspace{-3mm}
    \subfigure[WT]{
        \label{fig:exp2_WT}
        \includegraphics[width=0.245\textwidth]{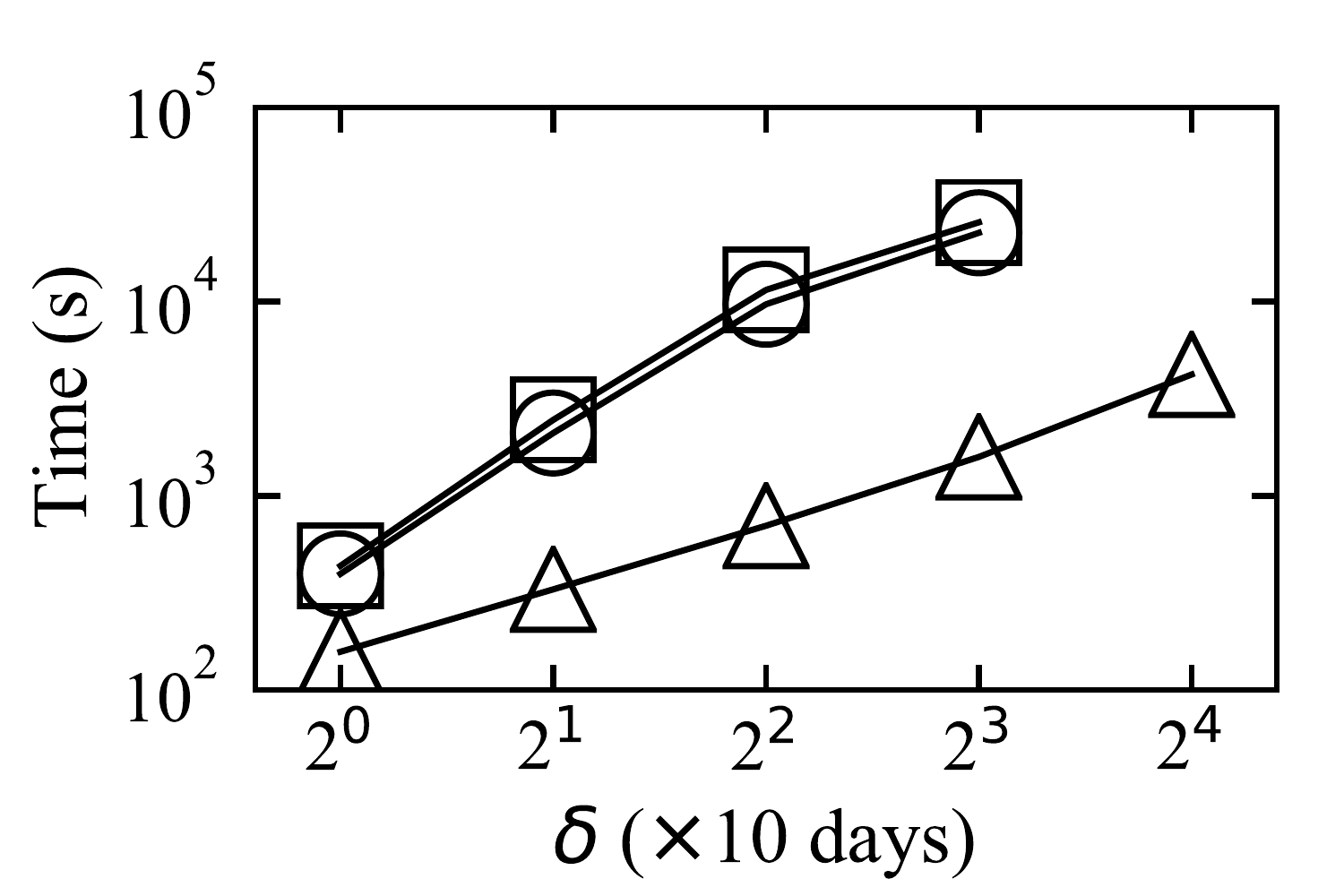}
    }\hspace{-3mm}
    \vspace*{-0.2in}
    \caption{Time on varying $\delta$.}
    \label{fig:exp2_time}
    \vspace*{-0.08in}
\end{figure*}

\begin{figure*}[t]
    \centering
    \vspace*{-0.17in}
    \begin{minipage}[t]{0.48\linewidth}
        \centering
        \subfigure[WN]{
            \label{fig:exp2_mem_WN}
            \includegraphics[width=0.495\textwidth]{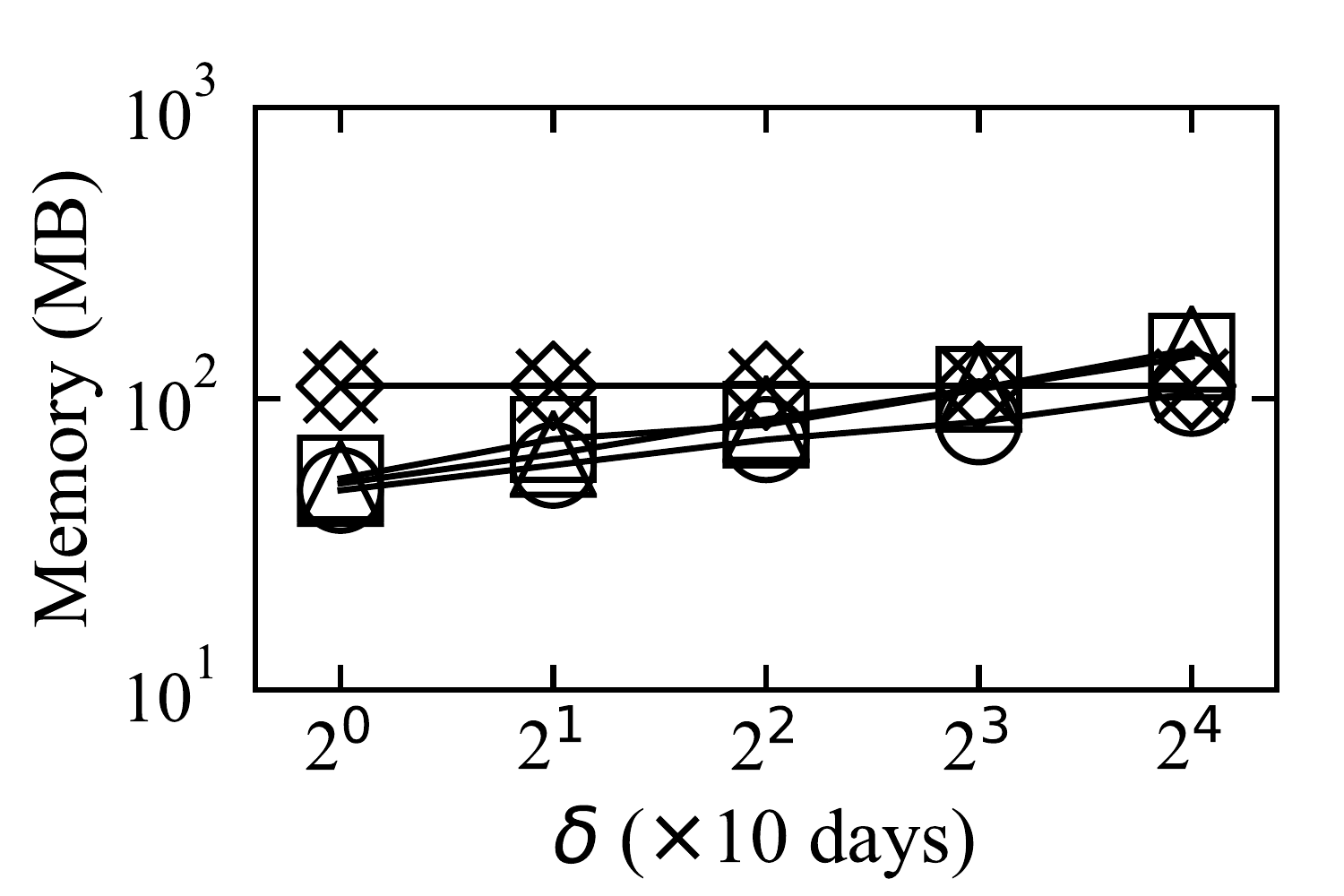}
        }\hspace{-3mm}
        \subfigure[ER]{
            \label{fig:exp2_mem_ER}
            \includegraphics[width=0.495\textwidth]{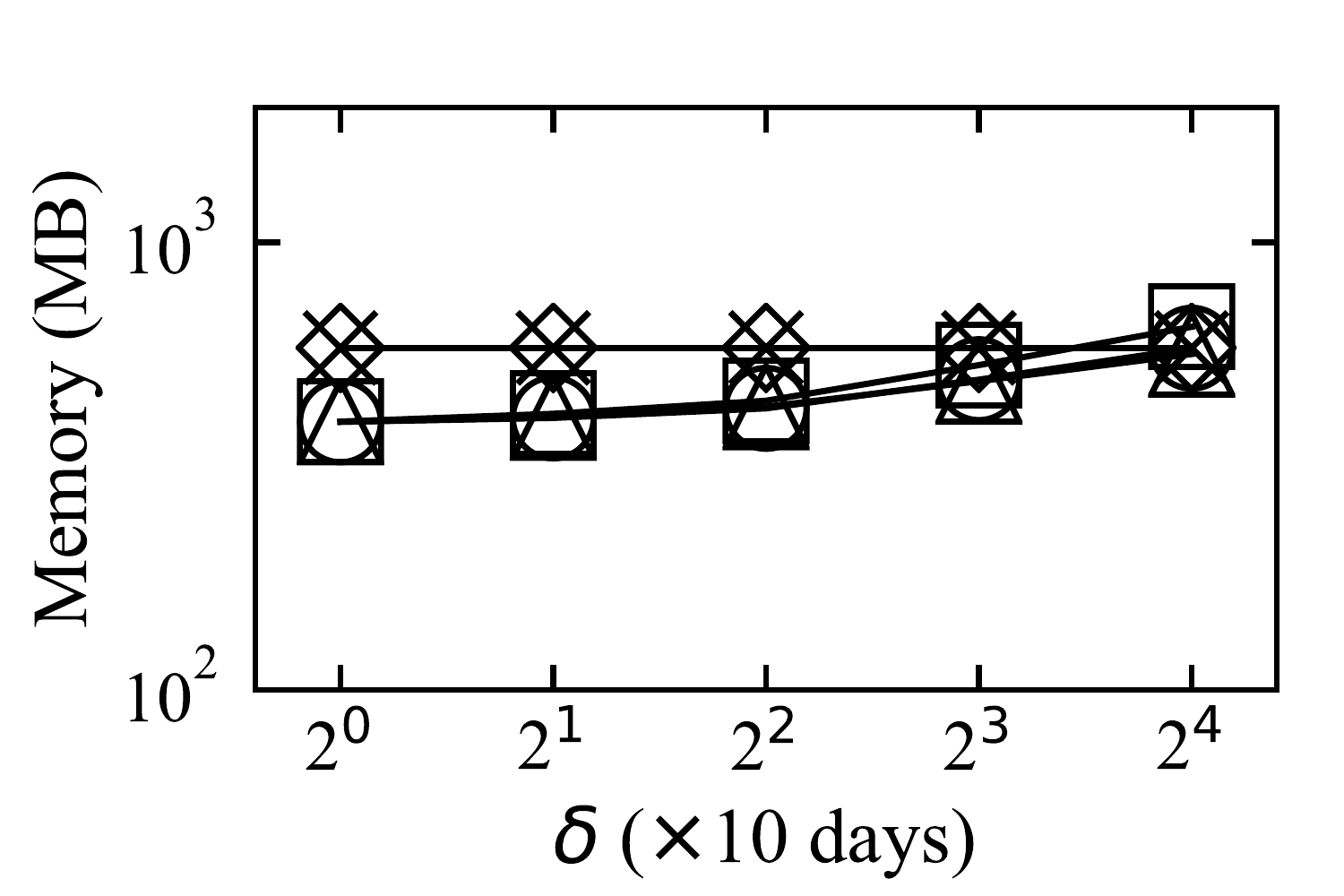}
        }\hspace{-3mm}
        \vspace*{-0.2in}
        \caption{Memory on varying $\delta$.}
        \label{fig:exp2_mem}
    \end{minipage}%
    \begin{minipage}[t]{0.48\linewidth}
        \centering
        \subfigure[ER]{
            \label{fig:exp4_ER}
            \includegraphics[width=0.495\textwidth]{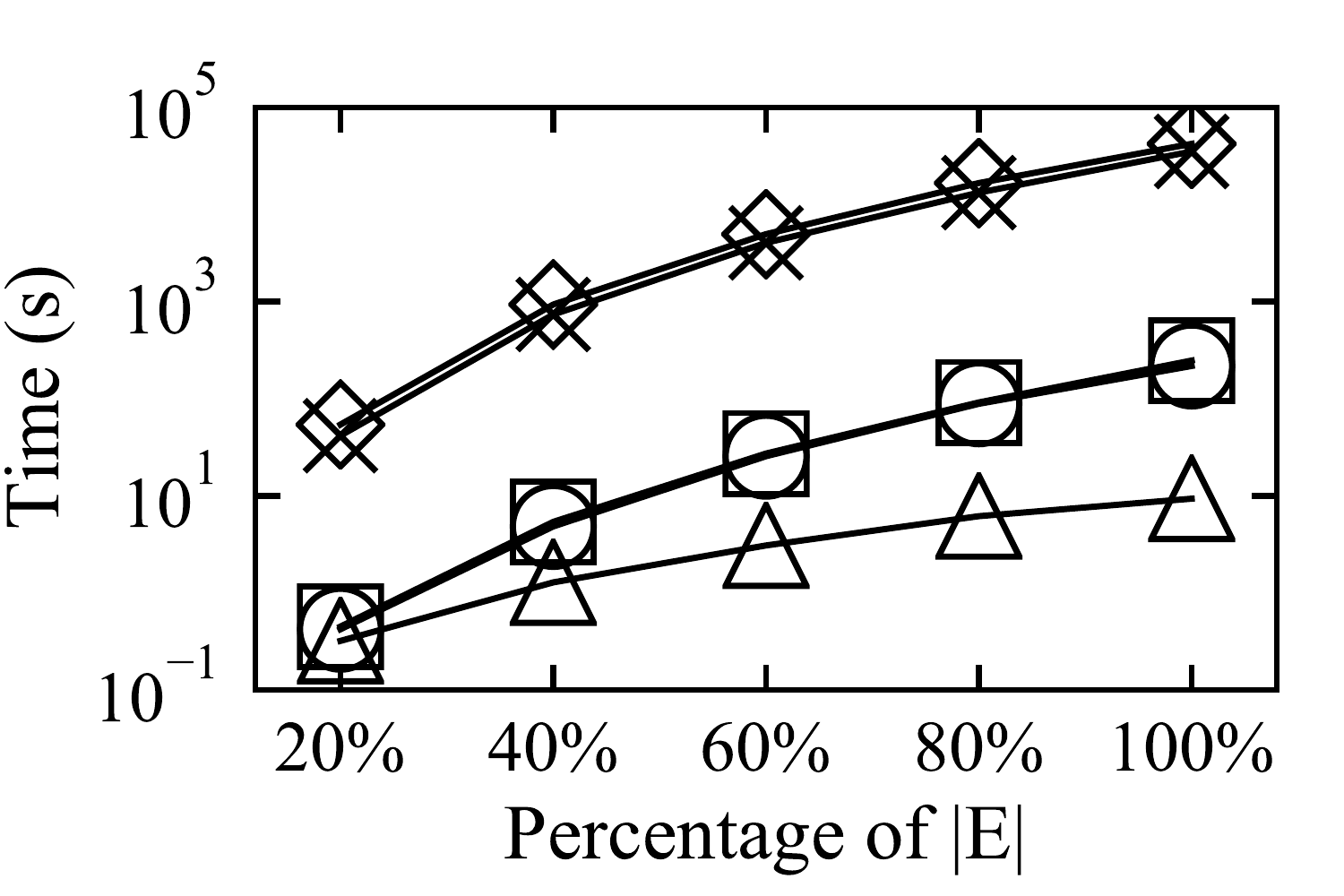}
        }\hspace{-3mm}
        \subfigure[WT]{
            \label{fig:exp4_WT}
            \includegraphics[width=0.495\textwidth]{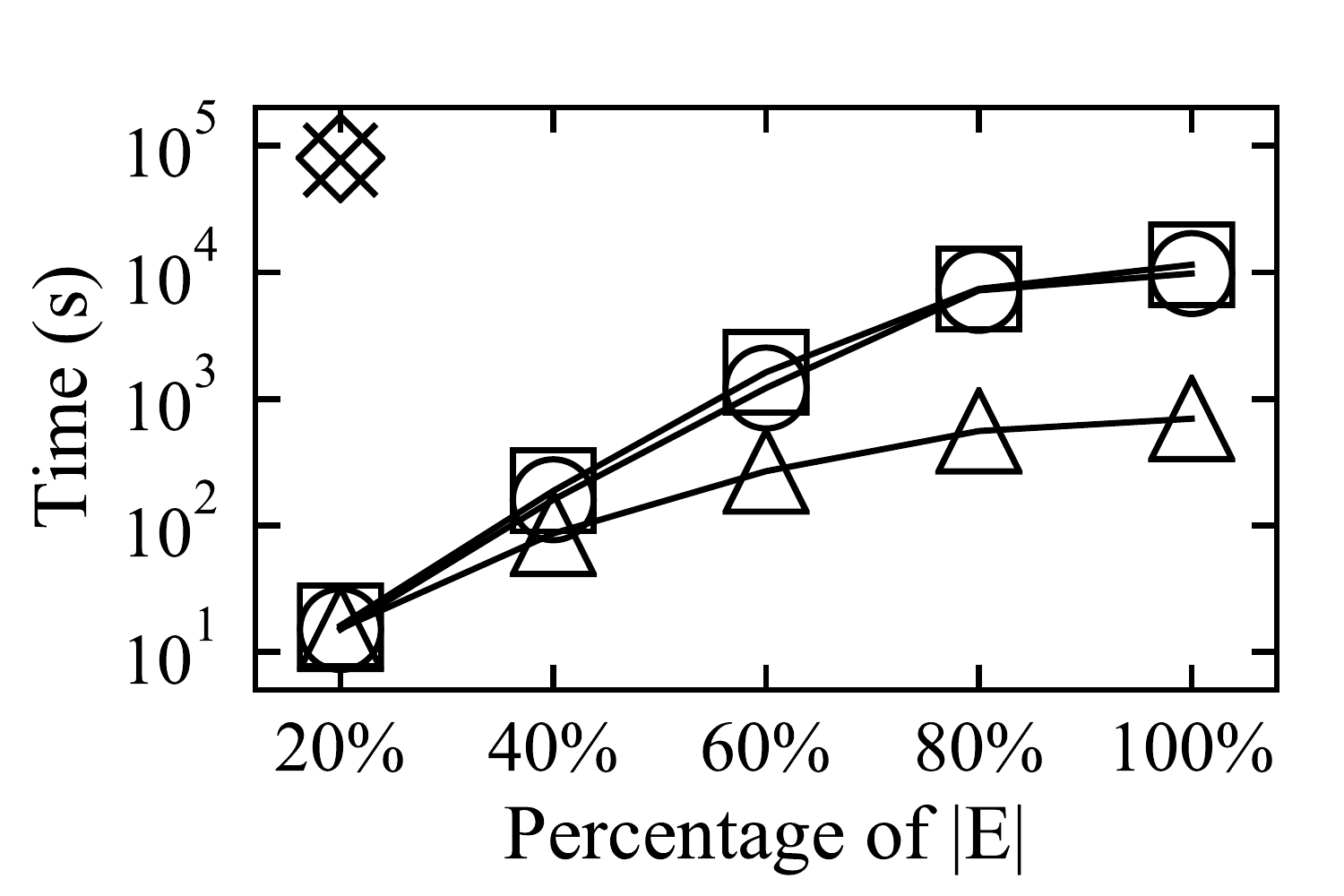}
        }\hspace{-3mm}
        \vspace*{-0.2in}
        \caption{Time on varying scale of $|E|$.}
        \label{fig:exp4_scal}
    \end{minipage}
    \vspace*{-0.15in}
\end{figure*}


As expected, \AlgoName{TBC} is the slowest counting algorithm and even exceeds the time limit on the LF and WT datasets. \rev{\AlgoName{TBC$^+$} demonstrates a speedup ranging from 1.9$\times$ to 161.9$\times$ compared to \AlgoName{TBC}. The performance of \AlgoName{TBC$^{++}$} is the most favorable, being comparable to \AlgoName{TBC$^+$} on certain datasets (e.g., SO, EP) and significantly outperforming it on others (e.g., WN, WT). For instance, on the WN dataset, \AlgoName{TBC$^{++}$} achieves a speedup of up to 61.3$\times$.} Notably, \AlgoName{TBC$^+$} and \AlgoName{TBC$^{++}$} perform similarly mostly on datasets that are ``easy to handle''. This is mainly due to the small size of the candidate wedge set in these datasets. Although \AlgoName{TBE} follows the same flow as \AlgoName{TBC}, it is slightly slower due to the additional time required for instance construction. Similarly, \AlgoName{TBE$^+$} performs similarly to \AlgoName{TBC$^+$} despite utilizing range traversal instead of binary search. Despite having the same theoretical complexity as \AlgoName{TBE}, the notable improvement of \AlgoName{TBE$^+$} over \AlgoName{TBE} provides strong evidence for the effectiveness of our proposed framework. The experimental results are consistent with our time complexity analysis, further validating our approach. Figure~\ref{fig:exp1_time} provides an illustration of the total counts of all six types of butterflies on different datasets, with the efficiency of the algorithms showing a positive correlation with the counts.

The memory cost is presented in Figure~\ref{fig:exp1_mem}. Clearly, memory consumption increases with the graph size. The memory consumption of the five algorithms is nearly identical, which aligns with our theoretical analysis. In some datasets (e.g., WN, ER), the optimization algorithms even exhibit lower memory overhead than the baseline algorithm, indicating the effectiveness of the pruning strategy during wedge enumeration. However, in certain datasets (e.g., TW, EP), the optimization algorithms incur slightly higher memory overhead due to the presence of auxiliary data structures. Nevertheless, the additional memory overhead of our optimization algorithms remains small compared to the significant efficiency improvements they provide. Furthermore, even on the largest ten-million-scale dataset WT, our algorithms require a minimal memory overhead of only 4GB.

\noindent\textbf{Effects of the Duration Constraint}. The duration constraint $\delta$ is the only parameter in our problem. Larger $\delta$ allows more temporal butterflies, resulting in a greater number of permutations.

\rev{As shown in Figure~\ref{fig:exp2_time}, among the algorithms tested, \AlgoName{TBC} and \AlgoName{TBE} perform the worst, while \AlgoName{TBC$^{++}$} shows the best performance. \AlgoName{TBC$^{+}$} and \AlgoName{TBE$^+$} fall in between.} As $\delta$ increases, the time cost of \AlgoName{TBC} and \AlgoName{TBE} remains nearly constant since they don't consider $\delta$ during the wedge enumeration process and explore every possible combination. On these ``easy to handle'' datasets (e.g., AM), the performance of optimization algorithms is identical. However, on other datasets, the time cost of \AlgoName{TBC$^+$} grows faster than \AlgoName{TBC$^{++}$}, which is reasonable because the wedge set grows as $\delta$ gets bigger (and subsequently more arrays in $HP$), and \AlgoName{TBC$^+$} will be affected greatly - it even runs out of time on WT dataset when $\delta=160$ days. The time cost gap between \AlgoName{TBE$^+$} and \AlgoName{TBC$^+$} widens as $\delta$ increases since the efficiency disparity between range traversal and the binary search becomes more noticeable with larger wedge sets.

The memory cost over varying $\delta$ is presented in Figure~\ref{fig:exp2_mem}. The baseline algorithms have a constant memory cost since they lack specific strategies for $\delta$. For small $\delta$, the optimization algorithms employ a pruning strategy during the wedge enumeration phase, resulting in a significant reduction in storage cost. However, as $\delta$ increases, the impact of the pruning strategy diminishes as fewer wedges can be filtered, and the auxiliary data structures' cost grows with the expanding wedge set. Consequently, the memory cost of the optimization algorithms increases continuously, possibly surpassing that of the baseline algorithms (see Figure~\ref{fig:exp2_mem_WN}). Nonetheless, the growth rate is much lower than that of $\delta$, and the overall memory overhead remains similar to that of the baseline algorithms.

\begin{figure*}[tbp]
    \begin{minipage}[t]{0.48\linewidth}
        \vspace*{-0.1in}
        \centering
        \subfigure[TW]{
            \label{fig:exp2_count_TW}
            \includegraphics[width=0.46\textwidth]{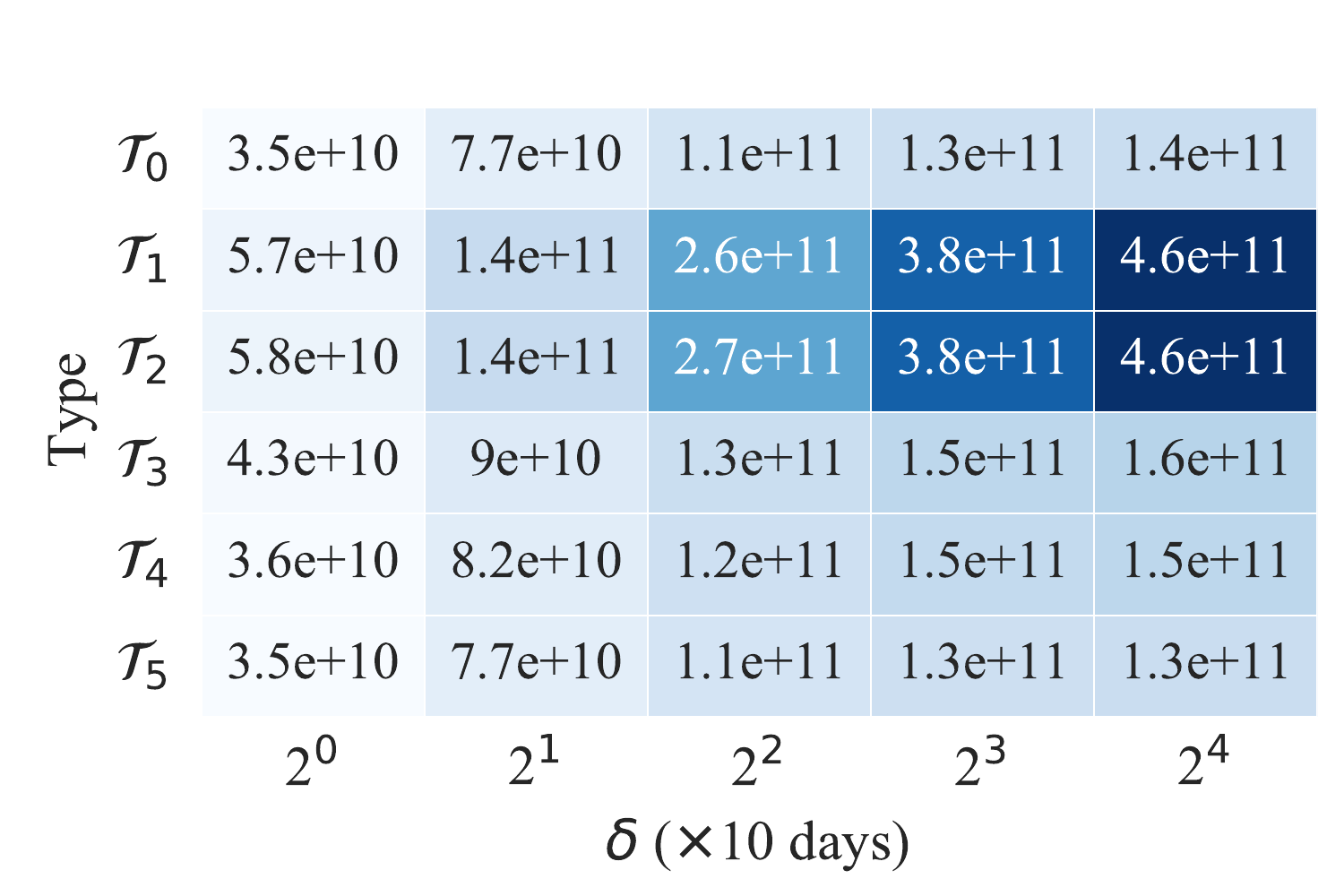}
        }\hspace{-3mm}
        \subfigure[EP]{
            \label{fig:exp2_count_EP}
            \includegraphics[width=0.46\textwidth]{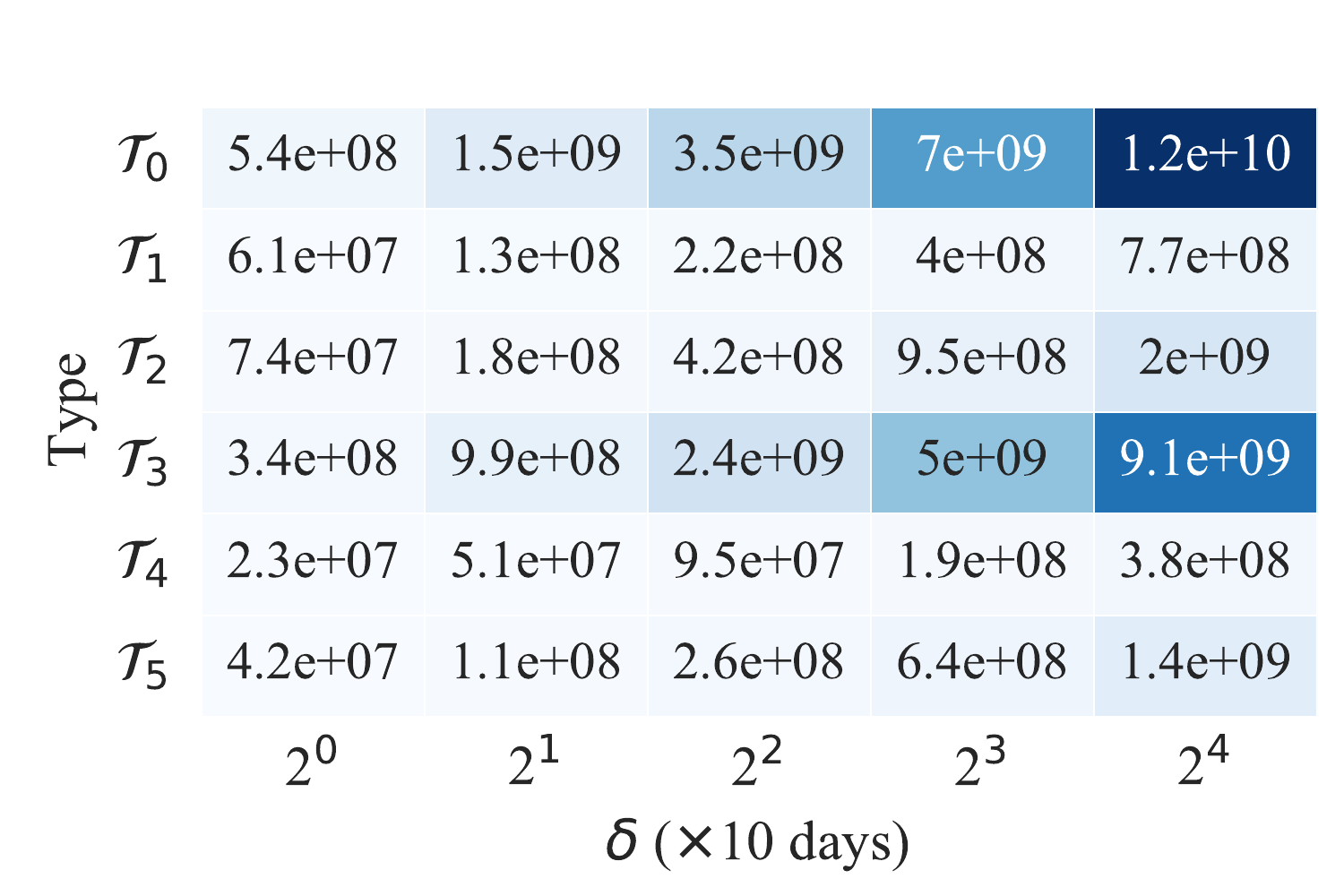}
        }\hspace{-3mm}
        \vspace*{-0.1in}
        \caption{Counts on varying $\delta$.}
        \label{fig:exp2_count}
        \includegraphics[width=\textwidth]{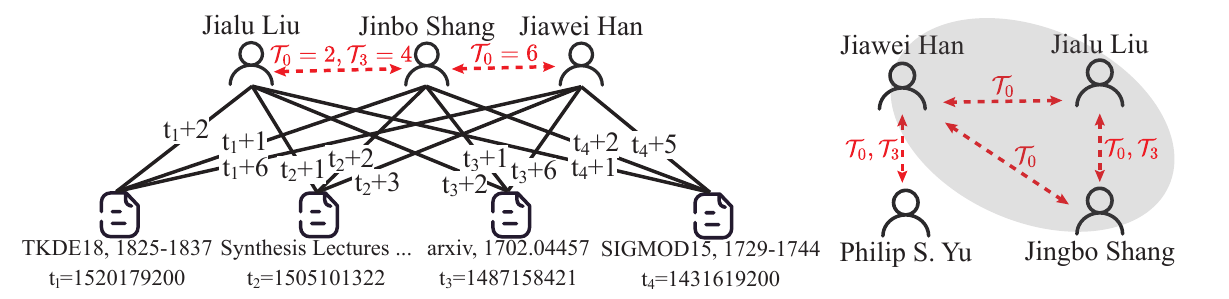}
        \vspace*{-0.3in}
        \caption{\rev{Case study on DBLP.}}
        \label{fig:casestudy}
        \vspace*{-0.3in}
    \end{minipage}%
    \hspace{2mm}
    \begin{minipage}[t]{0.50\linewidth}
        \centering
        \setlength{\tabcolsep}{2pt}
        \makeatletter\def\@captype{table}\makeatother
        \caption{The distribution of counts while $\delta=40$ days.}
        \vspace{2mm}
        \begin{tabular}{|c|c||c|c|c|c|c|c|}
            \hline
            \multirow{2}{*}{\textbf{Dataset}} & \multirow{2}{*}{\textbf{Entities}} & \multicolumn{6}{c|}{\textbf{Percentage of Total Counts}} \\ \cline{3-8}
            & & $\mathcal{T}_0$ & $\mathcal{T}_1$ & $\mathcal{T}_2$ & $\mathcal{T}_3$ & $\mathcal{T}_4$ & $\mathcal{T}_5$\\ \hline
            \cellcolor{lime}WQ & user-page & 18.4\% & \cellcolor{lightgray}\textbf{22.6\%} & \cellcolor{lightgray}\textbf{29.5\%} & 15.2\% & 6.9\% & 7.5\% \\ \hline
            \cellcolor{lime}ER & user-page & 17.1\% & \cellcolor{lightgray}\textbf{34.1\%} & \cellcolor{lightgray}\textbf{24.0\%} & 12.2\% & 7.2\% & 5.4\% \\ \hline
            \cellcolor{lime}WT & user-page & 15.8\% & \cellcolor{lightgray}\textbf{19.8\%} & \cellcolor{lightgray}\textbf{19.7\%} & 16.6\% & 14.3\% & 13.8\% \\ \hline
            TW & user-tag& 11.1\% & \cellcolor{lightgray}\textbf{26.2\%} & \cellcolor{lightgray}\textbf{26.3\%} & 13.1\% & 12.2\% & 11.0\% \\ \hline
            LF & user-band & 15.1\% & \cellcolor{lightgray}\textbf{21.6\%} & \cellcolor{lightgray}\textbf{21.8\%} & 16.9\% & 13.1\% & 11.6\% \\ \hline
            \cellcolor{pink}CU & tag-publication & \cellcolor{lightgray}\textbf{20.6\%} & 15.1\% & \cellcolor{lightgray}\textbf{19.7\%} & \cellcolor{lightgray}\textbf{20.6\%} & 11.3\% & 12.7\% \\ \hline
            \cellcolor{pink}BS & tag-publication & \cellcolor{lightgray}\textbf{21.0\%} & 13.0\% & \cellcolor{lightgray}\textbf{19.4\%} & \cellcolor{lightgray}\textbf{22.1\%} & 10.9\% & 13.6\% \\ \hline
            \cellcolor{orange}SO & user-post & \cellcolor{lightgray}\textbf{19.3\%} & \cellcolor{lightgray}\textbf{20.5\%} & \cellcolor{lightgray}\textbf{19.2\%} & \cellcolor{lightgray}\textbf{21.8\%} & 10.0\% & 9.2\% \\ \hline
            \cellcolor{orange}AM & user-product & \cellcolor{lightgray}\textbf{23.1\%} & \cellcolor{lightgray}\textbf{19.6\%} & \cellcolor{lightgray}\textbf{19.2\%} & \cellcolor{lightgray}\textbf{20.7\%} & 9.1\% & 8.4\% \\ \hline
            WN & user-page & \cellcolor{lightgray}\textbf{30.1\%} & 12.2\% & 12.6\% & \cellcolor{lightgray}\textbf{19.8\%} & \cellcolor{lightgray}\textbf{20.2\%} & 5.1\% \\ \hline
            EP & user-product & \cellcolor{lightgray}\textbf{51.1\%} & 3.2\% & 6.1\% & \cellcolor{lightgray}\textbf{34.4\%} & 1.4\% & 3.8\% \\ \hline
        \end{tabular}\label{tab:distribution}
    \end{minipage}
    \vspace*{-0.09in}
\end{figure*}

Figure~\ref{fig:exp2_count} displays the counts of each type of temporal butterflies with varying $\delta$, where darker grids represent higher counts. As $\delta$ increases, the counts also tend to rise. The rate of increase varies across different datasets and types of temporal butterflies, making it unpredictable. Furthermore, in line with previous studies~\cite{paranjape2017motifs, vazquez2004topological, yaverouglu2014revealing, milo2004superfamilies}, the distribution of temporal butterflies' counts shows minimal variation as the threshold changes. For instance, on the EP dataset, type $\mathcal{T}_0$ consistently accounts for approximately half the total, while type $\mathcal{T}_3$ always constitutes 30\% of the total counts.


\noindent\textbf{Distribution of Different Temporal Butterfly Types}. Table~\ref{tab:distribution} presents the count distribution, highlighting prevalent types. Clear distinctions are observed across various datasets, but commonalities also exist (datasets with the same cell color). In datasets like WQ, ER, and WT, where edges represent user-page edits, butterfly types $\mathcal{T}_1$ and $\mathcal{T}_2$ appear frequently, accounting for at least 39\% of the total count. This indicates the moderate follower effect (less significant $\mathcal{T}_0$) as page editing usually requires time. CU and BS datasets, where edges denote tag assignments in CiteULike and BibSonomy, show higher percentages of types $\mathcal{T}_0$, $\mathcal{T}_2$, and $\mathcal{T}_3$. In the EP dataset (Epinions' user-product rating network), types $\mathcal{T}_0$ and $\mathcal{T}_3$ constitute almost 85\% of the total count, with $\mathcal{T}_0$ alone making up over half. SO and AM datasets, representing marks between users and items (e.g., posts and products), exhibit a relatively balanced and prominent distribution of types $\mathcal{T}_0$, $\mathcal{T}_1$, $\mathcal{T}_2$, and $\mathcal{T}_3$. Notably, types $\mathcal{T}_4$ and $\mathcal{T}_5$ usually appear less frequently. In a nutshell, these observations suggest that focusing on some specific butterfly types, such as $\mathcal{T}_0$ for the strong follower effect, could lead to enhanced results while minimizing unnecessary effort.

\noindent\textbf{Scalability}.The scalability of the algorithms is depicted in Figure~\ref{fig:exp4_scal}, illustrating the running time of all competitors across different graph sizes. To evaluate scalability, we randomly select a portion of edges (i.e., $\{20\%, 40\%, 60\%, 80\%\}$) from the initial datasets, apply our method to the induced graph, and average the running time over 10 iterations. As anticipated, the time overhead for all algorithms increases with the percentage of edges, albeit at varying rates. Notably, \AlgoName{TBC$^+$}, \AlgoName{TBE$^+$}, and \AlgoName{TBC$^{++}$} exhibit excellent scalability, with computation costs increasing linearly relative to the percentage of edges. Among them, \AlgoName{TBC$^{++}$} demonstrates the best performance. On the WT dataset, our baseline algorithms only achieve completeness when the percentage is set to 20\%.

\noindent\rev{\textbf{Case Study}. We collect an author-paper dataset from DBLP\footnote{https://dblp.org/} using the search key ``mining'', comprising 50,536 papers and 78,459 distinct authors. To timestamp the edges, we use the paper publication time with an additional offset determined by author order as a tie-breaking rule.
Upon analysis, we uncovered an interesting trend: the closer the collaboration between authors, the more tense butterflies we observed. Most notably, the majority of butterflies in our dataset fell into $\mathcal{T}_0$ and $\mathcal{T}_3$, mainly due to the substantial time gap between publications and the assigned offset.
We extract a representative part of the data, as shown in Figure~\ref{fig:casestudy}. Here, we find that the number of butterflies between Jialu Liu and Jinbo Shang is evenly split between $\mathcal{T}_0$ and $\mathcal{T}_3$, suggesting a balanced collaboration between them. In contrast, when examining Jiawei Han in relation to the first two authors, all butterflies were $\mathcal{T}_0$. This pattern may indicate a guiding relationship within these collaborations. The left part of Figure~\ref{fig:casestudy} presents details about their collaborative papers.
Further investigation reveals that Jialu Liu and Jinbo Shang are graduate students closely collaborating, while Jiawei Han serves as their primary supervisor. Similarly, we can find that Jiawei Han and Philip S. Yu are scholars who work closely together, as indicated by the even distribution between $\mathcal{T}_0$ and $\mathcal{T}_3$.}

\begin{figure*}[tbp]
	\begin{minipage}[t]{0.76\linewidth}
		\centering
		\includegraphics[width=0.60\textwidth]{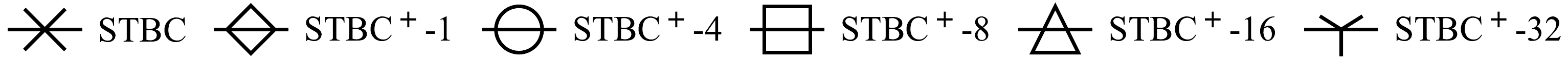}\\
		\vspace*{-0.21in}
		\begin{minipage}[t]{0.494\linewidth}
			\subfigure[LF]{
				\label{fig:exp3_wLF}
				\includegraphics[width=0.494\textwidth]{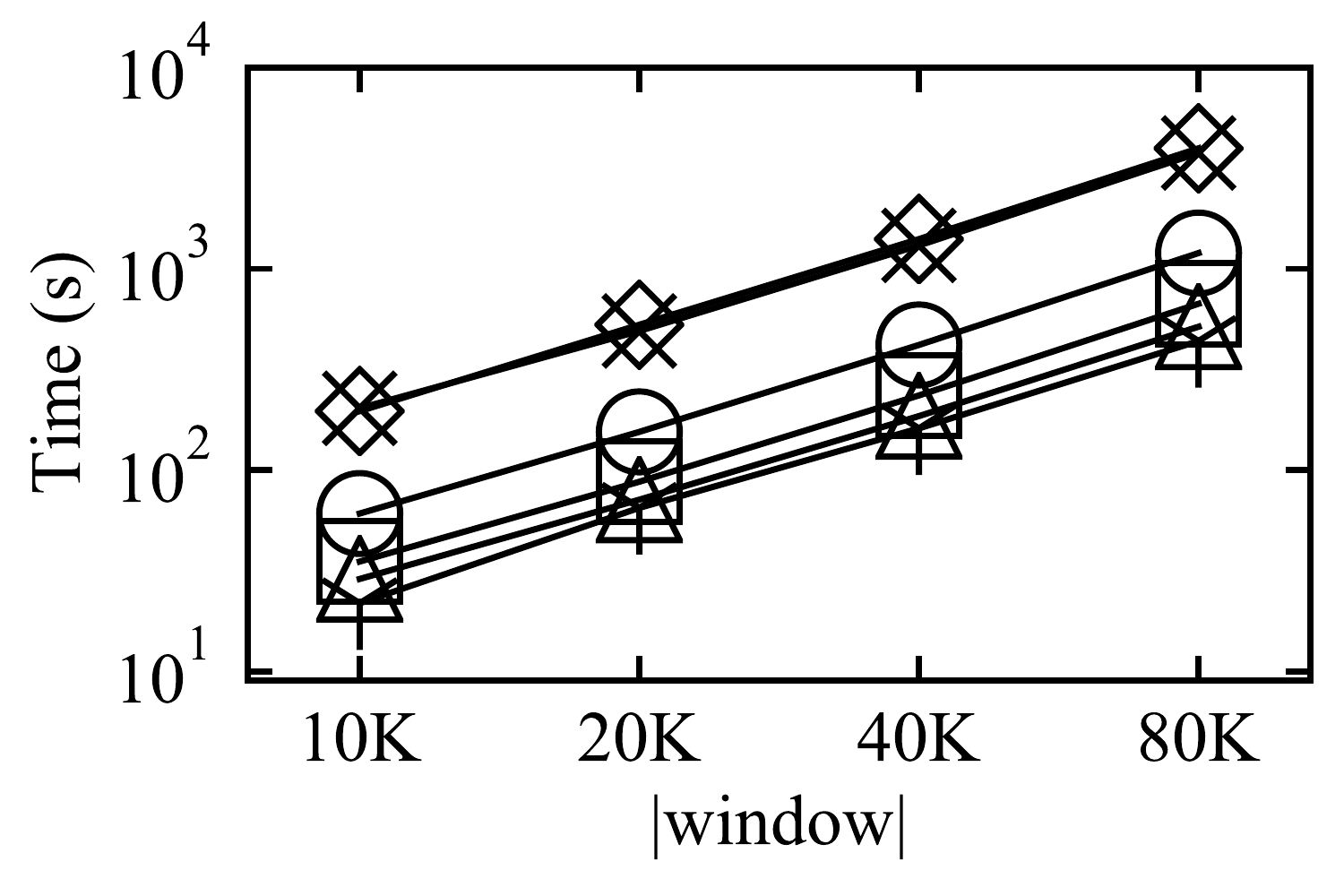}
			}\hspace{-3mm}
			\subfigure[WT]{
				\label{fig:exp3_wWT}
				\includegraphics[width=0.494\textwidth]{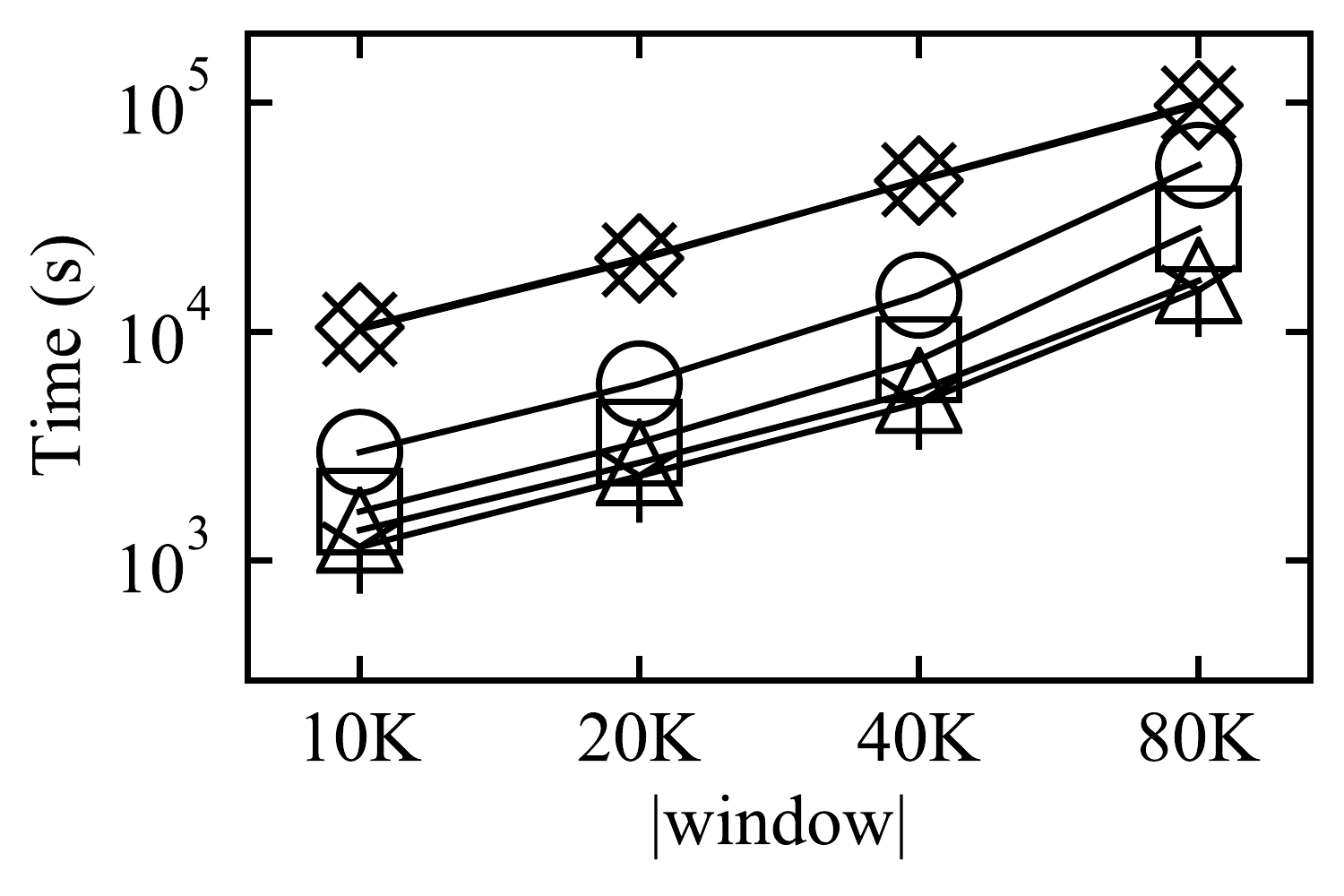}
			}\hspace{-3mm}
			\vspace*{-0.2in}
			\caption{\rev{Time on varying $|window|$.}}
			\label{fig:exp3_window}
		\end{minipage}
		\begin{minipage}[t]{0.494\linewidth}
			\subfigure[LF]{
				\label{fig:exp3_sLF}
				\includegraphics[width=0.494\textwidth]{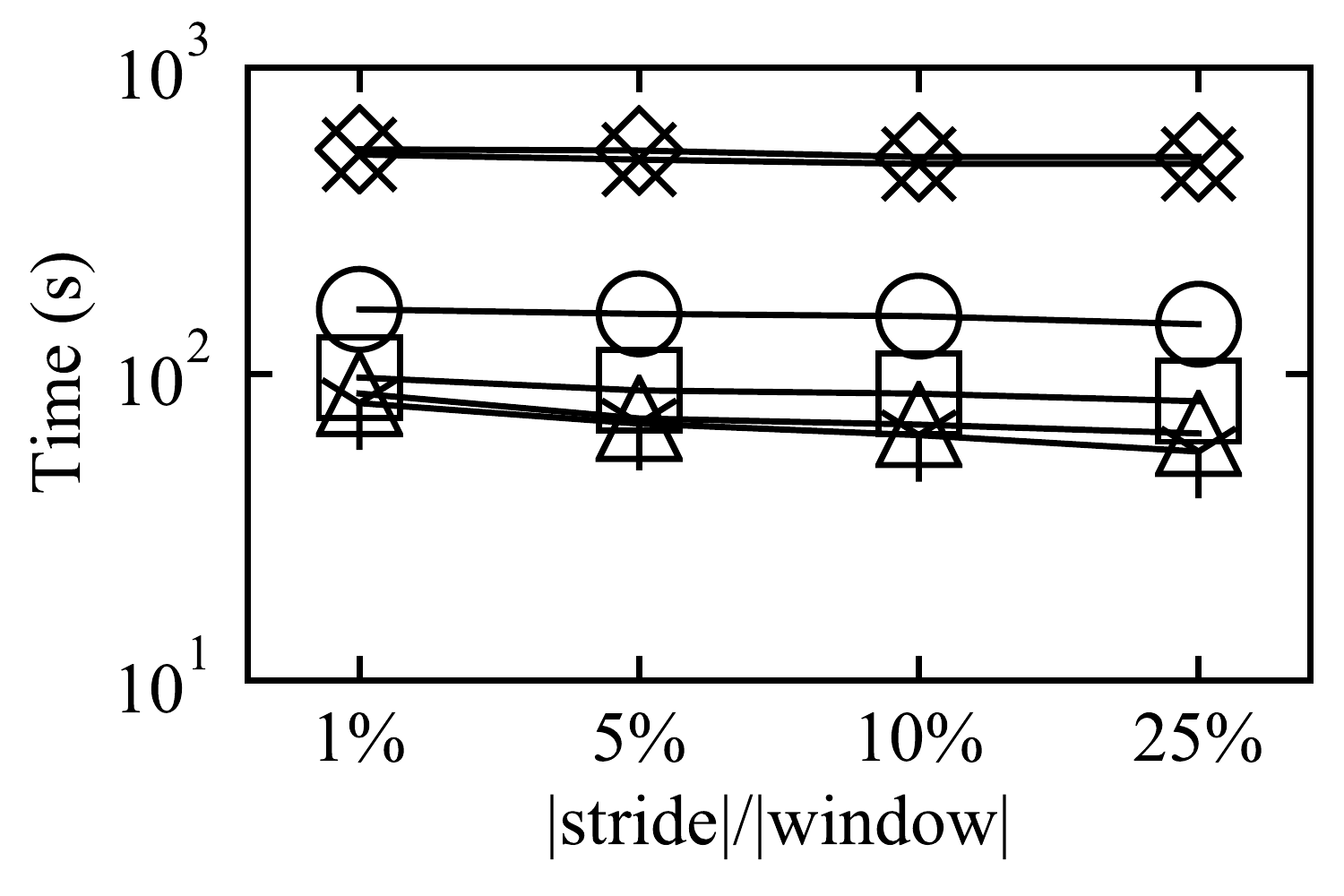}
			}\hspace{-3mm}
			\subfigure[WT]{
				\label{fig:exp3_sWT}
				\includegraphics[width=0.494\textwidth]{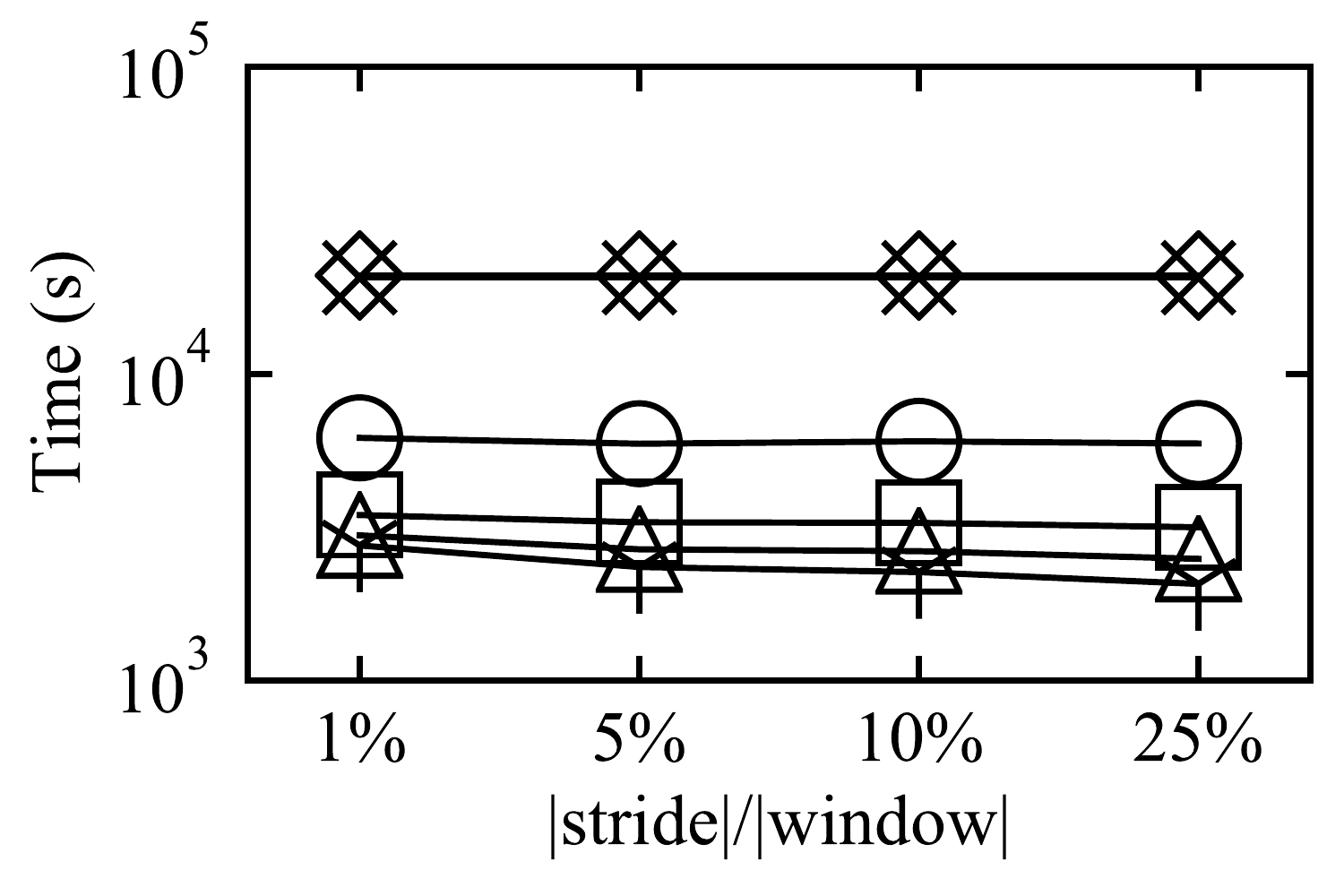}
			}\hspace{-3mm}
			\vspace*{-0.2in}
			\caption{\rev{Time on varying $|stride|$/$|window|$.}}
			\label{fig:exp3_stride}
		\end{minipage}
	\end{minipage}
	\begin{minipage}[t]{0.235\linewidth}
		\centering
		\includegraphics[width=0.998\textwidth]{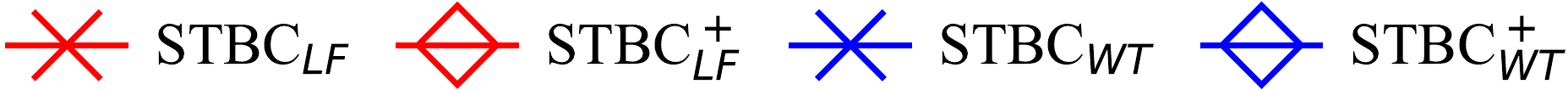}
		\includegraphics[width=0.77\textwidth]{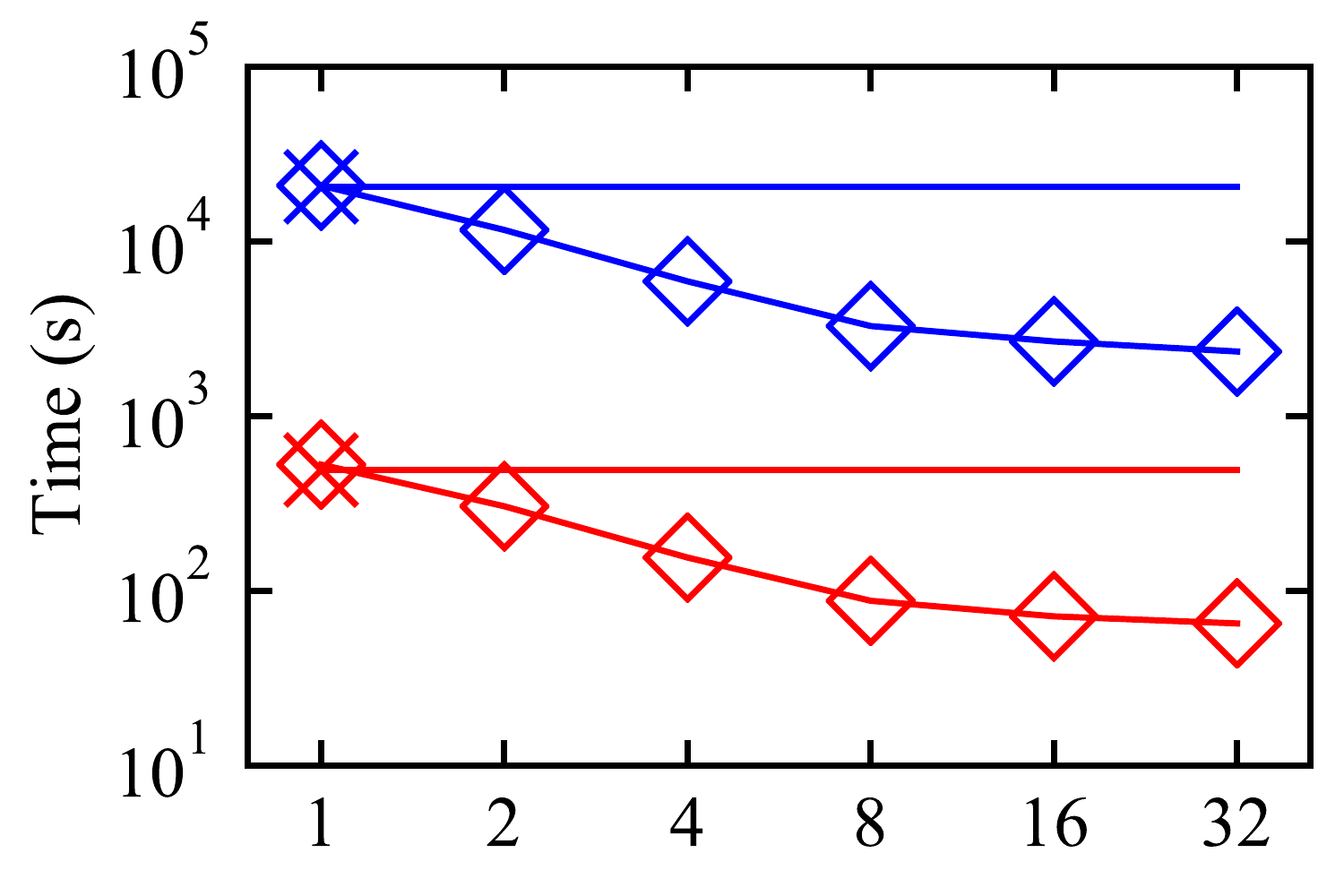}
		\vspace*{-0.1in}
		\caption{\rev{Time on varying $|thread|$.}}
		\label{fig:exp3_thread}
	\end{minipage}
	\vspace*{-0.2in}
\end{figure*}

\subsection{Evaluation on Graph Streams}
\noindent\textbf{Varying Window Sizes}. Figure~\ref{fig:exp3_window} showcases the efficiency of \AlgoName{STBC} and \AlgoName{STBC$^+$} for varying $|window|$ sizes, with a fixed $|stride|$ set to 5\% of $|window|$. Additionally, we evaluate the performance of \AlgoName{STBC$^+$} with different thread sizes, denoted as \AlgoName{STBC$^+$-4} for a 4-thread parallel algorithm, for example. As the $|window|$ size increases, the time cost of all algorithms also increases due to the larger candidate wedge set in the induced graph.

Comparing \AlgoName{STBC$^+$-1} and \AlgoName{STBC}, they exhibit similar efficiency since they have the same time complexity. \rev{However, \AlgoName{STBC$^+$} with multi-threading shows significantly improved speed, with \AlgoName{STBC$^+$-32} being up to 12.7 times faster than \AlgoName{STBC} on the WT dataset. It's worth noting that on the LF dataset, \AlgoName{STBC$^+$-1} is slower than \AlgoName{STBC} when $|window|$ is greater than or equal to 20K. This is because \AlgoName{STBC$^+$} needs to insert all edges before counting, resulting in an actual graph size of $|window|+|stride|$ during querying, which slows it down compared to \AlgoName{STBC}.} However, this drawback is tolerable because $|stride|$ is always much smaller than $|window|$, and the benefits of multi-threading far outweigh the cost.

\noindent\textbf{Varying Stride Sizes}. The evaluations in Figure~\ref{fig:exp3_stride} analyze the impact of varying $|stride|$ on algorithm performance, with a fixed window size of 20K. For \AlgoName{STBC}, stability against different stride values is observed due to its sequential edge updates and consistent graph size during querying. Initially, the time cost of \AlgoName{STBC$^+$} decreases as larger strides facilitate load balancing among threads. However, this improvement diminishes as load balancing approaches its limit. \rev{Notably, on the WT dataset, \AlgoName{STBC$^+$} experiences a slowdown when the stride percentage is $\ge 5\%$, indicating a slight impact on algorithm efficiency, consistent with findings from evaluations on varying window sizes.} In summary, when ample computational resources are available, \AlgoName{STBC$^+$} with multi-threading is the preferred choice.

\noindent\textbf{Varying Thread Sizes}. Figure~\ref{fig:exp3_thread} presents the results obtained with different thread sizes while keeping $|window|$ and $|stride|$ at their default settings. \rev{The evaluation includes \AlgoName{STBC$_{LF}$}, \AlgoName{STBC$^{+}_{WT}$}, \AlgoName{STBC$_{LF}$}, and \AlgoName{STBC$^{+}_{WT}$} for the LF and WT datasets, respectively.} It is important to note that \AlgoName{STBC} is limited to single-thread execution and cannot be parallelized. The results clearly demonstrate that \AlgoName{STBC$^+$} outperforms \AlgoName{STBC} significantly when multiple threads are utilized. However, the acceleration effect gradually diminishes due to resource limitations and the overhead associated with thread context switching. Further exploration of the trade-off between algorithm efficiency and resource cost holds promising potential for future investigation.

%% file: 7-RelatedWork.tex
\section{Related Work}
\label{sec:rel}


\noindent\textbf{Butterfly on Bipartite Graphs}. Significant research efforts have been dedicated to the study of  \ProbName{butterfly counting and enumeration}, which is the most fundamental sub-structure in bipartite graphs. Wang \textit{et al.}~\cite{wang2014rectangle} first propose the \ProbName{butterfly counting} problem and counting through enumerating wedges from a randomly selected layer. Sanei-Mehri \textit{et al.}~\cite{sanei2018butterfly} further develop a strategy for choosing the layer while Wang \textit{et al.}~\cite{WangLQZZ19} achieve state-of-the-art efficiency by employing vertex priority. \rev{Additionally, various techniques have been explored in \ProbName{butterfly counting}, including parallel processing~\cite{sanei2018butterfly, shi2022parallel}, external memory optimization~\cite{WangLQZZ19}, sampling~\cite{sanei2018butterfly, li2021approximately, sheshbolouki2022sgrapp}, GPU~\cite{xu2022efficient}, and batch update~\cite{wang2022accelerated}. Recent advancements have extended the \ProbName{butterfly counting} problem to bipartite graph streams~\cite{sanei2019fleet, sheshbolouki2022sgrapp} and uncertain bipartite graphs~\cite{zhou2021butterfly, zhou2023butterfly}.} Yang \textit{et al.}~\cite{yang2023p} propose a competitive search-based method for counting and enumerating the ($p,q$)-bicliques, with the butterfly serving as a special case where $p,q$=$2$.
\FullVersion{\rev{Temporal butterflies, unlike their static counterparts determined sorely by four vertices, carry richer information due to the presence of various temporal edges between vertex pairs. This complexity affects the validation of temporal butterflies and their real-world implications. Yang's algorithm \AlgoName{BCList$^{++}$}~\cite{yang2023p}, which relies on a vertex-centric approach and preprocessing 2-hop vertices, cannot seamlessly adapt to this temporal context.
Furthermore, even when disregarding the preprocessing time required by \AlgoName{BCList$^{++}$}, it is found that \AlgoName{BCList$^{++}$} only achieves marginally better efficiency than \AlgoName{BFC-VP}~\cite{WangLQZZ19} on smaller datasets but lags behind \AlgoName{BFC-VP} on the majority of datasets. Consequently, \AlgoName{BFC-VP} is recognized as the prevailing state-of-the-art butterfly counting algorithm, and it is our chosen foundation for extending our baseline algorithm.}}
It is noteworthy that although the majority of research on butterflies has predominantly concentrated on counting, most of these methods can be readily expanded to facilitate enumeration with minimal modifications (\cite{sanei2018butterfly, WangLQZZ19,zhou2021butterfly}).

\noindent\textbf{Temporal Motif on Temporal Unipartite Graphs}. The problem of \ProbName{temporal motif counting and enumeration} holds particular significance as the butterfly motif represents a distinct type of motif within this context. Building upon the concept of \ProbName{motif counting}~\cite{ribeiro2014discovering, watts1998collective}, \ProbName{temporal motif counting} has been extensively studied recently~\cite{liu2021temporal, boekhout2019efficiently, li2018temporal, sun2019new, wang2020efficient, gurukar2015commit}, but their definitions vary. Kovanen \textit{et al.}~\cite{kovanen2011temporal} introduce the concept of $\triangle T$-adjacency, which pertains to two temporal edges sharing a vertex and having a timestamp difference of at most $\triangle T$. Additionally, they take into account the temporal ordering aspect. Redmond \textit{et al.}~\cite{redmond2013temporal} study the $\delta$-temporal motif counting without temporal ordering. The most relevant work to ours, Paranjape \textit{et al.}~\cite{paranjape2017motifs} define $\delta$-temporal motif where edges in the motif are within $\delta$ duration and the temporal ordering is considered as well. Pashanasangi \textit{et al.}~\cite{pashanasangi2021faster} introduce different thresholds for the time difference between each pair of adjacent edges in a temporal triangle. While numerous studies have focused on the 3-vertex temporal motif counting~\cite{kovanen2011temporal, paranjape2017motifs, gao2022scalable, pashanasangi2021faster}, \rev{Boekhout \textit{et al.}~\cite{boekhout2019efficiently} delve into the 4-vertex temporal motifs, but specifically omitted the discussion of temporal rectangle. They propose that there are 3 non-isomorphic temporal rectangles on the unipartite graph and require a specially designed solution. Fortunately, our research encompasses this problem as a subset, and our techniques readily handle it.} Furthermore, there are numerous approximation algorithms available for solving counting problems~\cite{liu2019sampling, sarpe2021oden}. When it comes to enumeration problems, isomorphism-based algorithms are the most commonly used~\cite{li2019time, mackey2018chronological, locicero2021temporalri}, but they lack optimizations tailored to specific motifs, resulting in low efficiency.

%% file: 8-Conclusion.tex
\section{Conclusion}
\label{sec:con}

\rev{In this paper, we investigate the \ProbName{temporal butterfly counting and enumeration} problem. We formally define the problem and propose a solution based on the state-of-the-art \ProbName{butterfly counting} algorithm. We further devise three optimization algorithms, two for counting and one for enumeration. Within a unified framework, these algorithms harness a combination of techniques, resulting in a significant reduction in overall time complexity without compromising on space efficiency.} Additionally, we extend our algorithms to address the practical scenario of graph streams and further propose a parallel algorithm. Finally, extensive experiments validate the efficiency and scalability of our proposed algorithms.

%% file: 9-Approximation.tex
\newpage
\appendix

\section{Study on Approximation}

In this section, we demonstrate that our algorithms can be seamlessly integrated into state-of-the-art approximate techniques to support approximate \ProbName{temporal butterfly counting}.

\noindent\textbf{Extension of the Approximation}. The state-of-the-art approximate algorithm for \ProbName{static butterfly counting}, \AlgoName{ApproxBFC} (Algorithm 8 in \cite{sanei2018butterfly}), uses a specified probability $p$ to decide whether to include each edge in a new graph. It returns the \textit{exact butterfly count} on this new graph multiplied by $p^{-4}$ as the {\em approximate butterfly count}.

For \ProbName{static butterfly counting on graph streams}, the leading algorithm, \AlgoName{sGrapp}~\cite{sheshbolouki2022sgrapp}, segments the stream into non-overlapping windows based on a parameter $N_t^W$. Each window contains exactly $N_t^W$ unique timestamps (except for the last one). It calculates butterfly counts within the window using an {\em exact counting algorithm} and then adds the estimated counts between windows to obtain the final result. Specifically, the approximate counts between each newly divided window and all previous windows is calculated by $EC^{\alpha}$, where $EC$ represents the total number of edges up to the current window, and $\alpha$ is a predetermined parameter that requires dynamic adjustments for varying data.

To adapt these approaches to \ProbName{temporal butterfly counting}, we seamlessly replace the exact \ProbName{static butterfly counting} part in these algorithms with our \ProbName{temporal butterfly counting} algorithms (\AlgoName{TBC}, \AlgoName{TBC$^+$}, \AlgoName{TBC$^{++}$}) and apply these approximate strategies to each type of the temporal butterfly.
This yields approximate algorithms, denoted as \AlgoName{ApproxTBC}, \AlgoName{ApproxTBC$^+$}, and \AlgoName{ApproxTBC$^{++}$}, as well as \AlgoName{sGrappTBC}, \AlgoName{sGrappTBC$^+$}, and \AlgoName{sGrappTBC$^{++}$}.
Notably, when adapting \AlgoName{sGrapp} (referring to Section~5 of~\cite{sheshbolouki2022sgrapp}), we need to empirically preset a unique $\theta$ parameter for each type, denoted as $\{\theta_i\}_{i=0}^5$. 

\noindent\textbf{Correctness}. The correctness of our approximation extensions aligns with the \ProbName{uncertain butterfly counting} algorithm in \cite{zhou2023butterfly}.
Particularly, it demonstrated that $\mathbb{E}[\hat{C_s}] = C_s$, where $C_s$ denotes the exact static butterfly count and $\hat{C_s}$ is the approximate count obtained by \AlgoName{ApproxBFC}~\cite{sanei2018butterfly}. By extending this to each type of temporal butterfly and excluding butterflies that violate temporal duration constraint, we establish $\forall_{i=0}^{5}\mathbb{E}[\hat{C}[i]]=C[i]$, where $C[i]$ denotes the exact temporal butterfly counts and $\hat{C}[i]$ denotes the approximate result of \AlgoName{ApproxTBC}, \AlgoName{ApproxTBC$^+$}, or \AlgoName{ApproxTBC$^{++}$}. In addition, the approximate strategy of \AlgoName{sGrapp} is based on the finding that the number of static butterflies changes exponentially with the graph stream, which is also applicable to each individual temporal butterfly type.


\noindent\textbf{Experimental Setup}. Building upon our earlier experimental analyses, we focused on two representative datasets, WN and TW, and conducted experiments for our approximate algorithms under identical settings. Each algorithm was executed ten times in every experiment with different seeds. Following~\cite{sheshbolouki2022sgrapp, sanei2019fleet}, we assess the accuracy of each algorithm using the average of mean absolute percentage errors (MAPE) for six temporal butterfly estimates, i.e., $\frac{1}{6}\sum_{i=0}^{5}\frac{|\hat{C}[i]-C[i]|}{C[i]}$, where $\hat{C}[i]$ represents the approximate counts and $C[i]$ signifies the true counts. The source code for these algorithms is openly available in our extended repository: https://github.com/ZJU-DAILY/TBFC-Extension.

\begin{figure}[htbp]
\vspace*{-0.1in}
    \centering
    \includegraphics[width=0.365\textwidth]{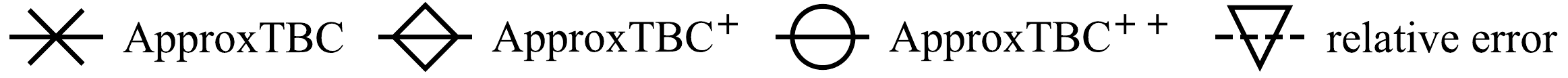}\\
    \vspace*{-0.06in}
    \subfigure[WN]{
        \label{fig:exp5_WN}
        \includegraphics[width=0.225\textwidth]{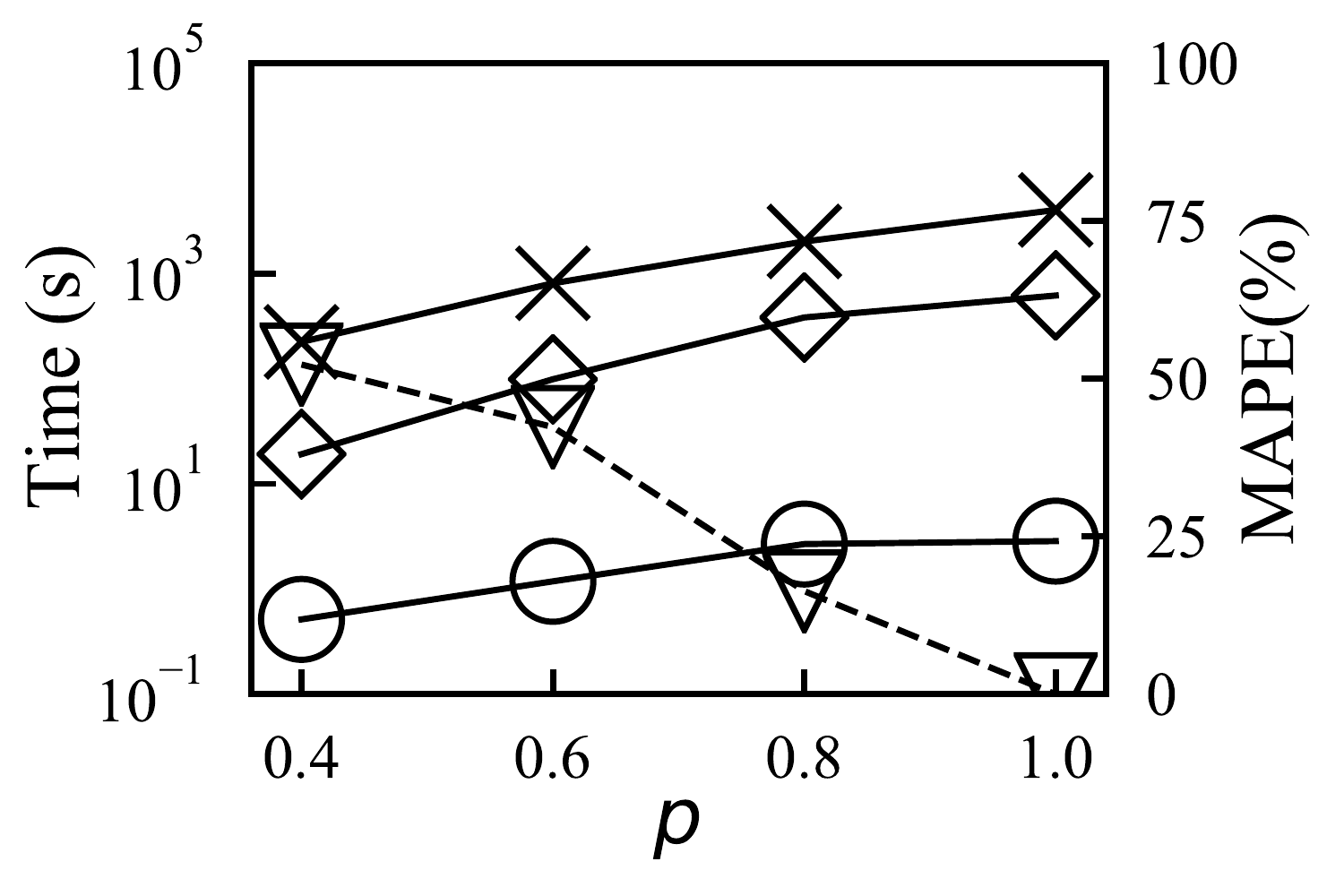}
    }
    \subfigure[TW]{
        \label{fig:exp5_TW}
        \includegraphics[width=0.225\textwidth]{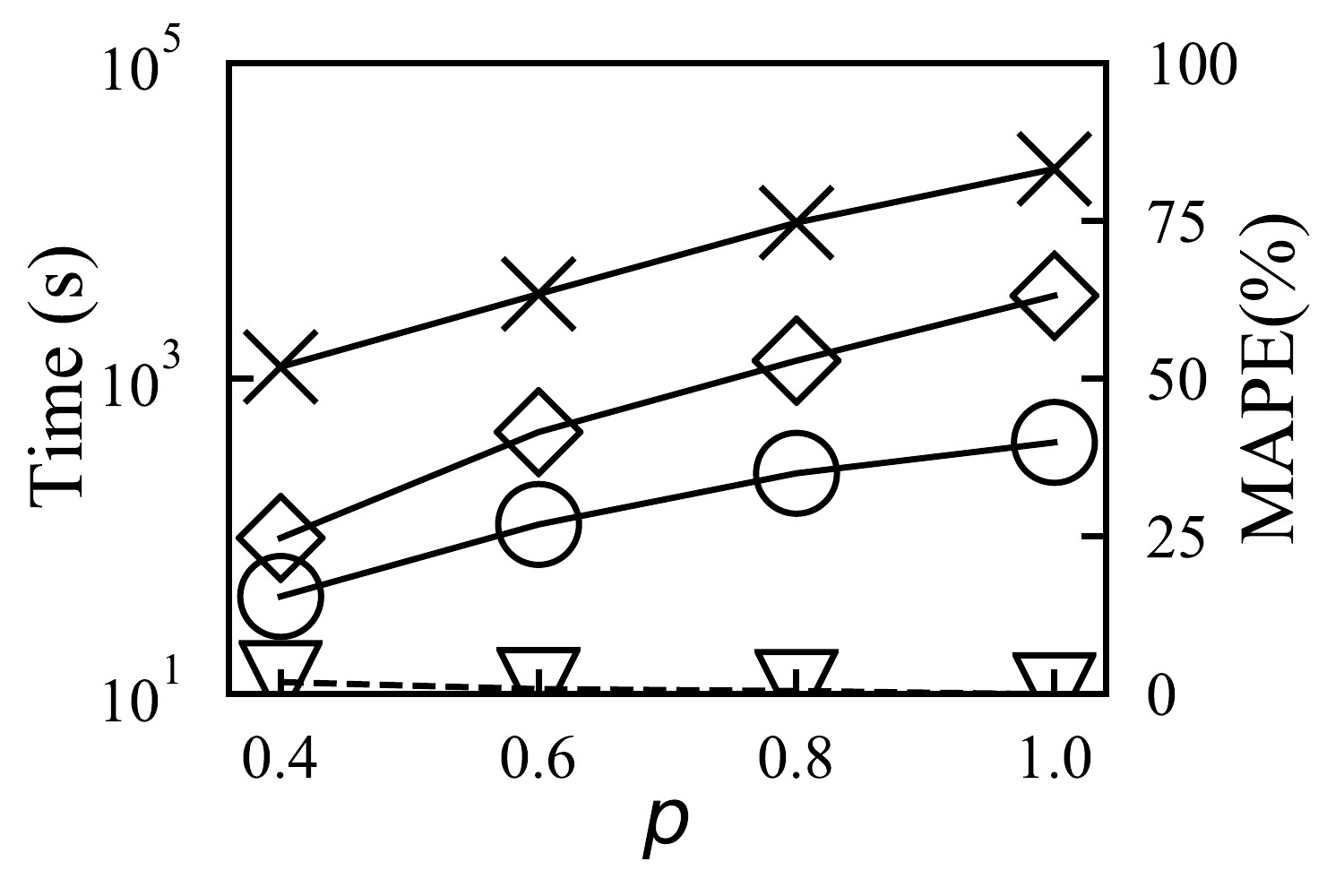}
    }
    \vspace*{-0.2in}
    \caption{Time on varying $p$.}
    \label{fig:exp5_Approx}
    \vspace*{-0.1in}
\end{figure}

\noindent\textbf{Experimental Evaluation}. In Figure~\ref{fig:exp5_Approx}, we present the time cost and relative error of our series of \AlgoName{Approx} algorithms across different probability thresholds ($p$). As anticipated, the time overhead for all algorithms increases with $p$, albeit at varying rates. Notably, \AlgoName{ApproxTBC$^+$}, and \AlgoName{ApproxTBC$^{++}$} exhibit excellent performance, with computation costs increasing linearly. Among them, \AlgoName{ApproxTBC$^{++}$} demonstrates the best performance, consistent with our experimental analysis of scalability. In terms of relative error, it's observed that as the parameter $p$ increases, the relative error tends to decrease. However, the algorithm's performance varies significantly across different datasets. For instance, on the WN dataset, the MAPE remains relatively high at 16.1\% even when $p=0.8$. In contrast, on the TW dataset, the error drops considerably to 1.9\% when $p=0.4$. Based on this finding, for small datasets, we may require a larger $p$ to sample more edges, thus improving our ability to capture the macroscopic features of the graph.

\begin{figure}[htbp]
\vspace*{-0.13in}
    \centering
    \includegraphics[width=0.365\textwidth]{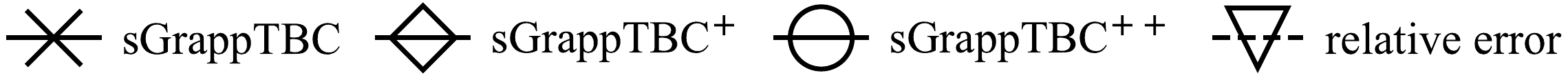}\\
    \vspace*{-0.06in}
    \subfigure[WN]{
        \label{fig:exp6_WN}
        \includegraphics[width=0.225\textwidth]{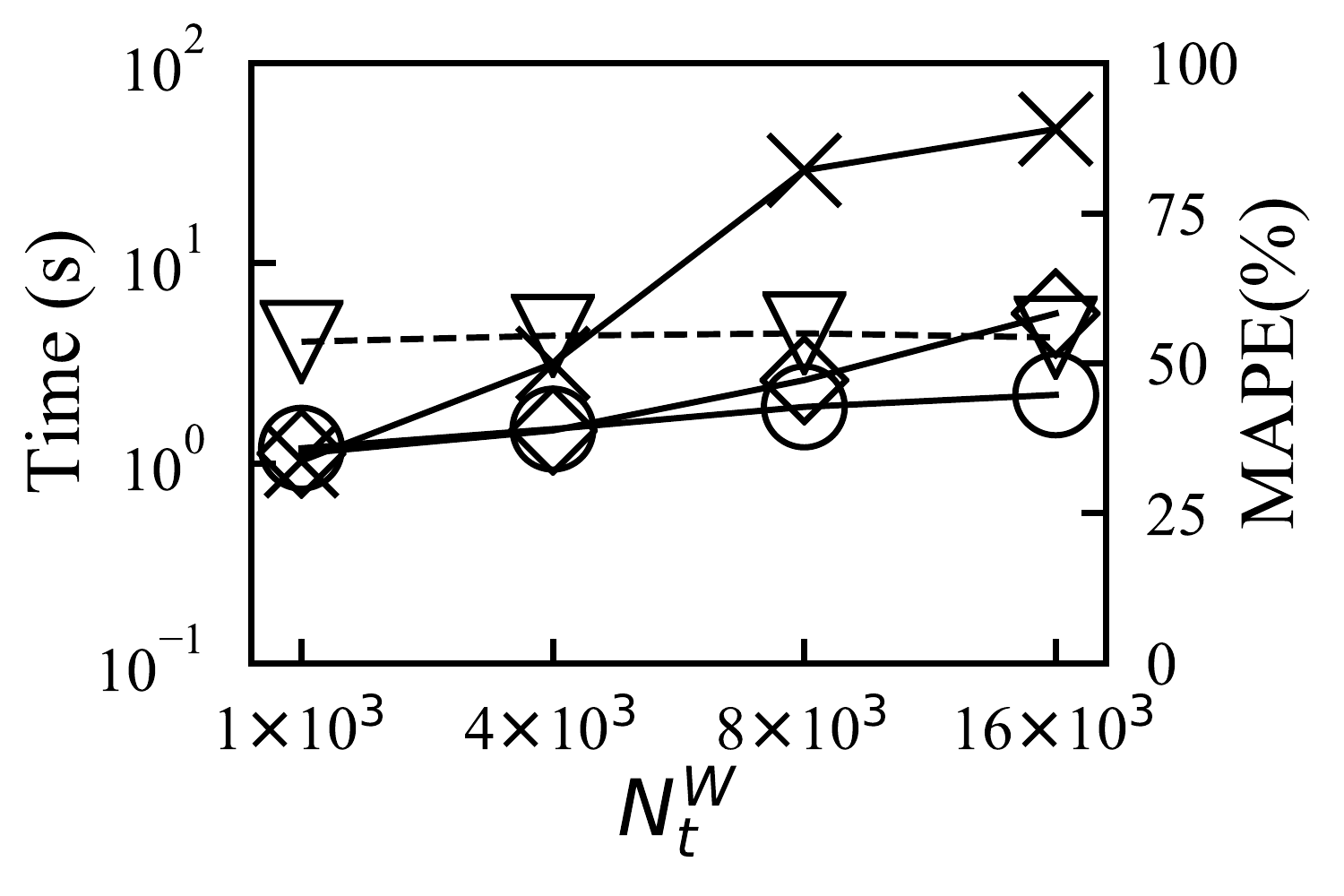}
    }
    \subfigure[TW]{
        \label{fig:exp6_TW}
        \includegraphics[width=0.225\textwidth]{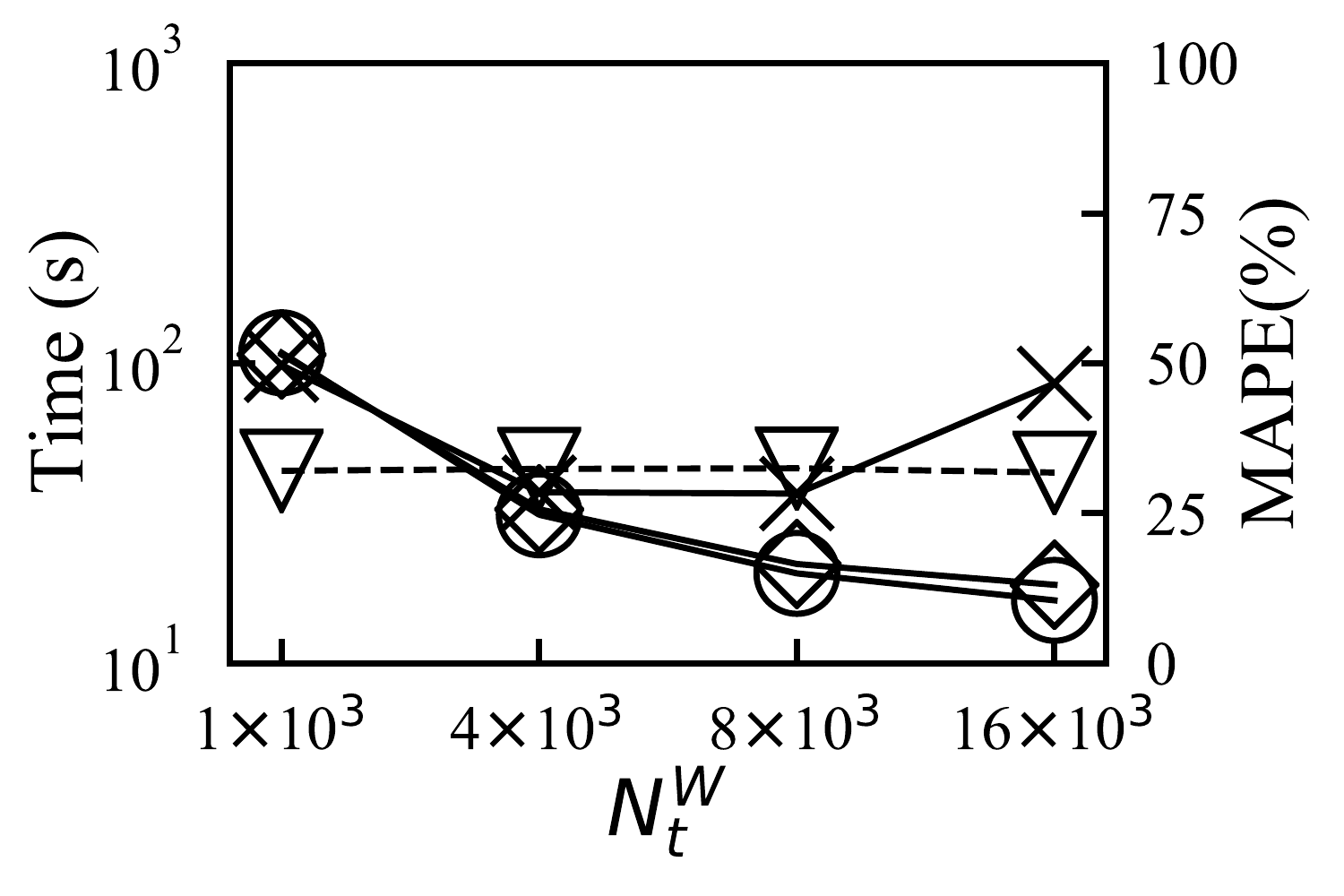}
    }
    \vspace*{-0.2in}
    \caption{Time on varying $N_t^W$.}
    \label{fig:exp6_sGrapp}
    \vspace*{-0.12in}
\end{figure}

The time cost and relative error of our series of \AlgoName{sGrapp} algorithms are depicted in Figure~\ref{fig:exp6_sGrapp}.
The time cost of these three algorithms based on sGrapp primarily depends on the parameter $N_t^W$. A larger $N_t^W$ means a larger single window and a smaller number of windows, making the choice of this parameter a trade-off as variations in the number and size of windows impact the algorithm's performance.	
On the WN dataset, all three algorithms show an increase in time overhead as $N_t^W$ increases. However, on the TW dataset, \AlgoName{sGrappTBC$^{+}$} and \AlgoName{sGrappTBC$^{++}$} decrease in time overhead as $N_t^W$ increases, while \AlgoName{sGrappTBC} exhibits a typical pattern, initially decreasing and then increasing. In general, \AlgoName{sGrappTBC$^{+}$} and \AlgoName{sGrappTBC$^{++}$} perform better and are less sensitive to the $N_t^W$ parameter.
To achieve optimal results, it's necessary to adjust the $\{\theta_i\}_{i=0}^5$ for different datasets and different $N_t^W$, typically within the range [1.0, 1.5]. The best relative error tends to remain relatively constant for a single dataset, with an average of 54.6\% for the WN dataset and 32.1\% for the TW dataset. As depicted by experiments in~\cite{sanei2018butterfly, sheshbolouki2022sgrapp}, the accuracy of \ProbName{static butterfly counting} approximate algorithms still has significant room for improvement in challenging graph streams. This holds true for our extended approximate algorithm as well and further highlights the importance of our proposed accurate algorithms, \AlgoName{STBC} and \AlgoName{STBC$^+$}, which address this issue and provide reliable results.